\documentclass[11pt]{article}
\usepackage[english]{babel}
\usepackage[T1]{fontenc}
\usepackage[]{amsfonts}
\usepackage{graphicx}
\newcommand{\ket}[1] {\mbox{$ \vert #1 \rangle $}}
\newcommand{\bra}[1] {\mbox{$ \langle #1 \vert $}}

\newcommand{\scal}[2] {\mbox{$ \langle #1 \vert #2 \rangle $}} 

\newcommand{\beq}{\begin{equation}}
\newcommand{\be}{\begin{equation}}
\newcommand{\eeq}{\end{equation}}
\newcommand{\ee}{\end{equation}}
\newcommand{\bea}{\begin{eqnarray}}
\newcommand{\eea}{\end{eqnarray}}

\renewcommand{\r}{\rho}

\renewcommand{\k}{\kappa}

\begin{document}
\begin{center}{\Large \bf Hawking Radiation from Acoustic Black Holes,  \\
Short Distance and Back-Reaction Effects}
\end{center}
\begin{center}
R. Balbinot$^{a}$,
A. Fabbri$^{b}$,
S. Fagnocchi$^{c,a}$\
and R. Parentani$^d$\
\end{center}
\footnotesize \noindent {\it a) Dipartimento di Fisica
dell'Universit\`a di Bologna and INFN sezione di Bologna, \\
\noindent via Irnerio, 46 - 40126 Bologna. E-mail: balbinot@bo.infn.it}\\
\noindent {\it b) Departamento de F\'{\i}sica Te\'orica and IFIC,
Facultad de F\'{\i}sica,
Universidad de Valencia, 46100 Burjassot-Valencia (Spain).  E-mail: afabbri@ific.uv.es}\\
\noindent {\it c)
Centro Enrico Fermi, Compendio Viminale, 00184 Roma (Italy). \\ E-mail: fagnocchi@bo.infn.it}\\
\noindent
{\it d)  Laboratoire de Physique Th\'eorique, CNRS UMR 8627,  
B\^atiment 210, Universit\'e Paris XI, 91405 Orsay Cedex (France).
E-mail: renaud.parentani@th.u-psud.fr}

\begin{abstract}

Using the action principle we first review how linear density
perturbations (sound waves) in an Eulerian fluid obey a
relativistic equation: the d'Alembert equation. This analogy
between propagation of sound and that of a massless scalar field
in a Lorentzian metric also applies to non-homogeneous flows. In
these cases, sound waves effectively propagate in a curved
four-dimensional ''acoustic'' metric whose properties are
determined by the flow. Using this analogy, we consider regular
flows which become supersonic, and show that the acoustic metric
behaves like that of a black hole. The analogy is so good that,
when considering quantum mechanics, acoustic black holes should
produce a thermal flux of Hawking phonons.

We then focus on two interesting questions related to Hawking
radiation which are not fully understood in the context of
gravitational black holes due to the lack of a theory of quantum
gravity. The first concerns the calculation of the modifications
of Hawking radiation which are induced by dispersive effects at
short distances, i.e., approaching the atomic scale when
considering sound. We generalize existing treatments and calculate
the modifications caused by the propagation near the black hole
horizon. The second question concerns backreaction effects. We
return to the Eulerian action, compute
 second order effects, and show that the backreaction of sound waves
 on the fluid's flow can be expressed in terms of their stress-energy tensor.
 Using this result in the context of Hawking radiation,
 we compute the secular effect on the background flow.
\end{abstract}

\section{Introduction and summary}
According to General Relativity black holes are spacetime regions
where the gravitational field is so strong that even light
is unable to escape.
It is indeed against all expectations-- including his own--
that Hawking
discovered  in 1974 \cite{hawking, hawking1} that black holes
are no longer black when Quantum Mechanics is taken into account.
More precisely he showed that
they 
{\it thermally} radiate (photons, neutrinos, gravitons, ...) with a temperature
inversely proportional to their mass. 
This 
discovery  had a major impact on theoretical physics. 

On a general level,
Hawking radiation (HR in the following)
finds its origin in the combination of the three
pillars of physics as presently understood, namely Special Relativity
(with the speed of light $c$ as fundamental constant), Quantum Field Theory
(the relativistic version of Quantum Mechanics
bringing $\hbar$), and General Relativity (the metric and relativistic
theory of gravity bringing Newton 
constant $G$).
This  triple
origin can be seen in the expression for the
temperature of a
(uncharged and unrotating)
black hole of mass $M$
\be
\k_B T_H = {1 \over 2 \pi} {\hbar c^3 \over 4 G M}\, ,
\ee
where $\k_B$ is Boltzmann constant. 

An experimental detection of HR would therefore be of great importance.
However, for a solar mass Black Hole
(BH in the following),
the temperature is only $T_H \simeq 6\times 10^{-8}\, K$.
This is much below
the Cosmic Microwave Background temperature $\simeq 2.7 \, K$. Hence there is no hope to detect
HR from astronomical BHs which result from a gravitational collapse,
since these should be more massive than the Chandrasekar mass
($\sim 3M_{sun}$) \cite{mtw}.

Nevertheless, there is still hope to detect HR.
First,
light BHs could have been produced in the early universe \cite{carr,musco}.
If their mass is smaller than $10^{15}gr$, their size would be of the order of one Fermi, and
they will have  
enough time to explode 
in the present epoch
at the end of the evaporation process
caused by the HR. The associated burst of radiation would possess
specific properties which could be detected in cosmic radiation \cite{barrau, PHB}.
Unfortunately it is improbable to find such BHs if the universe underwent a period of inflation,
as presently indicated by cosmological data (see for example \cite{WMAPinterpretation}).
Indeed, inflation would dilute
them away like any other pre-existing population of objects. Moreover
there seems to be no efficient mechanism to produce them
after inflation has stopped.
Finally, at present, there is no observational evidence for
excesses of radiation which could be interpreted as the bursts of radiation
emitted during the last stage of the
evaporation process. Nevertheless, 
the research in observational data is
still currently active, see e.g. \cite{search}.

Perhaps a more promising experimental possibility of detecting HR, albeit indirect,
arises from the astonishing analogy 
between light propagation in a curved spacetime and sound
propagation in a (non-uniformly) moving fluid \cite{unruh81},
or more generally, wave propagation in non-homogeneous condensed matter
systems \cite{volovik}.
Indeed this 
analogy works so well that acoustic Black Holes
(
i.e. the region
where sound is trapped)
should emit a thermal flux of phonons for the same reasons that
gravitational BHs emit HR.
This analogy is the central topic of this review article.
Its precise content will be 
explained later.
For the moment
we further discuss the various ramifications of HR
because this presentation 
illustrates the usefulness of the analogy. 

On one hand, HR solved the paradoxical aspects related to the
proposal \cite{bek} made
by Bekenstein in 1972 according to which black holes should possess an entropy proportional to the horizon area. Indeed the
properties of Hawking Radiation 
confirm that black holes
obey the laws of Thermodynamics and those of Statistical Mechanics
(in fact not only BH possess thermodynamical properties but all
event horizons in General Relativity, see e.g. \cite{horizonentropy}).
In a sense, HR is a necessity dictated by the second principle of thermodynamics
in the presence of horizons. We mention this aspect because it provides
a first indication that HR should not rely on the details of the
underlying fundamental theory, since its existence can be inferred by
a thermodynamical argument.

On the other hand,
HR has raised,
and it is still raising, new concepts and new questions
which concern intriguing
(and poorly understood) aspects of space time dynamics, such as the microscopical
origin of thermodynamics of horizons in the presence of gravity.
In fact, the very existence of HR 
poses deep questions concerning Quantum Gravity and the validity of
its semi-classical treatment based on quantum fields propagating
in a classical fixed background metric. It is this treatment
which has been so far used to derive HR (see however
\cite{CallanMalda, KeskiVakkuri, masspar}).
Among the 
questions raised by HR, two shall be discussed in this article,
in the light of the analogy with condensed matter systems.
The first one is related to the occurrence of very high frequencies.
This question is
generally referred to as the trans-Planckian problem, and it has played an
important role in the development of the 
above mentioned analogy \cite{ted91,unruh95,bmps}.
The second issue concerns the backreaction effects, i.e. the
consequences
on the background metric
induced by the non-linearities of the dynamical equations \cite{corto, lungo}.

\subsection{ Topics}

We 
first give
the basis of the kinematical analogy
between gravitational 
BH and sonic BH.
Then we 
explain why an acoustic BH should also emit HR,
and how the
 trans-Planckian question is addressed
in the acoustic framework, knowing that
the notion of acoustic wave ceases to exist when reaching the atomic scale.
Finally we discuss  the
backreaction effects due to HR in the acoustic setting.
Let us add that these 
last two sections should be considered as
two first steps of a more ambitious program: to understand
how to formulate backreaction effects at a deeper level
when the microscopic structure of the fluid 
(or that of space-time) is taken into account.


\subsubsection{Near horizon propagation and Hawking Radiation}

It should be first stressed that  both
HR and the occurrence of very high frequencies
result from the kinematical properties governing field propagation
near
a horizon.
Therefore these effects are not tied to BHs, nor even to General Relativity,
but to near horizon propagation in general.
When propagating backwards in time the field configurations giving rise
to HR, there is a blue shift effect which grows without bound as the horizon is approached.
Explicitly one finds
\be
\Omega(r) \propto {\omega \over 1 - r_h/r} \, ,
\label{boundlessg}
\ee
where $r-r_h$ is the distance from the horizon,
$\omega$ the asymptotic frequency, and $\Omega(r)$
the frequency measured by freely falling observers crossing the
horizon.
This boundless growth results from the absence of a Ultra-Violet 
(UV) scale which could stop it.
This absence is itself a direct consequence of Relativity: when studying fields obeying
a relativistic field equation,
their propagation in the near horizon geometry is governed by a
simplified and universal equation
which is {\it scale invariant},  the mass and the spin of the field having dropped out.
This universal behavior guarantees that all fields are subject to 
HR, but also, commitantly, that no UV scale could possibly stop
the blue-shifting effect of Eq. (\ref{boundlessg}).

This boundless growth is 
kinematical in the sense that it
follows from the properties of outgoing null geodesics, which are the characteristics of the relativistic
field equation. These characteristics hug the horizon, on both sides, according to
\begin{equation}
\label{expon}
x= r - r_h= x_0 \,
e^{k (t - t_0)/c} \, ,
\end{equation}
where
$k$ is the so-called
surface gravity of the horizon,
 and $t$ is a time coordinate
simply related to proper time at rest and far away from the hole.
For a 
black hole of mass $M$, $k=c^4/4GM = c^2/2 r_h$
(for an acoustic black hole,
$k$ is 
governed by the gradient of the velocity of the fluid
evaluated at the horizon 
and $t$ is the laboratory time). As we shall see in the text, this
exponential behavior will be associated, both for acoustic and
gravitational BHs, to the logarithmic singularity on the horizon
of stationary modes. Moreover, in both cases, the regularity of
the quantum state across the horizon implies the existence of HR
with a temperature \be
\k_B T_H= {\hbar k \over 2\pi c} \, , 
\label{univr}\ee
where $c$ is the speed of light for gravitational BHs or the speed of sound for acoustic BHs.
The important message is that there is a universal relationship
between the characteristics obeying Eq. (\ref{expon}),
the singular behavior of stationary modes subject to Eq. (\ref{boundlessg}),
and the existence of HR with a temperature given by Eq. (\ref{univr})
as a manifestation of regularity of the 
quantum state across the horizon.


\subsubsection{ The trans-Planckian question}

The boundless growth of Eq. (\ref{boundlessg})
is questionable \cite{ted91,ted93} both for acoustic and
gravitational BHs. In the second case, this is 
because it has been obtained
from quantum (test) fields propagating in a classical (passive) 
gravitational background.
In other words, Eq. (\ref{boundlessg}) 
 follows from the implicit assumption
that the quantum dynamics of gravity can be neglected, an {\it a priori}
good approximation
for large BHs for which the curvature and the temperature are far
from the Planck scale.\footnote{The Planck curvature is defined as
$l_P^{-2}$, where $l_P=\sqrt{G\hbar/c^3}=1.616\times 10^{-33}cm$,
and the Planck temperature as $E_P/\k_B$, where $E_P=\sqrt{\hbar
c^5/G}=1.221\times 10^{19}GeV$. } However, given the boundless
growth, this approximation is questionable since the Planck scale
is crossed in a logarithmically short time in the unit of
$(k/c)^{-1}=10^{-5} sec$ for a solar mass
BH.\footnote{ One might question this statement since one can
think that one can avoid large frequencies by a judicious choice
of coordinates. However this is not the case: the gravitational
tree-level scattering amplitude between an outgoing wave and an
infalling one
 is a scalar, and grows boundlessly when approaching the horizon
as $\Omega(r)$ 
does it in Eq. (\ref{boundlessg}) \cite{thooft, parentaPeyresq}.}
 Therefore it is not unreasonable to assume that quantum
gravitational effects will prevent the boundless growth of
frequencies, thus invalidating the semi-classical treatment.
However, if this aspect of the physics is not correctly described
by this treatment, shouldn't we also question the existence and
the properties of Hawking radiation, which have been derived using
these approximate settings ?

In the absence of a manageable theory of Quantum Gravity,
no systematic analysis can be performed to settle this question (see \cite{thooft, thooft2, parentaPeyresq}
for attempts and for references).
It is in this rather hopeless situation that the analogy with
phonon propagation near an acoustic horizon is useful.
The reason for this is rather simple. At low frequency, the analogy between
photon propagation in a given gravitational background
and phonon propagation in an acoustic metric is perfect: in both cases
the quantum field obeys the d'Alembert equation.
Therefore, if the acoustic metric possesses an event horizon, one expects to
find a thermal emission of Hawking phonons.
However, when reaching the molecular scale $d_c $
(a short distance with respect to the characteristic 
length scale $\kappa^{-1}
=(k/c^2)^{-1}$),
phonon propagation is modified and no longer described by the d'Alembert equation.
Indeed, in fluids, there is a UV regime (for wavelenghts smaller than $ d_c$),
in which the phonons
are described by a non-linear dispersion relation, which breaks the
Lorentz symmetry which emerges 
at low frequency.

\subsubsection{Two lessons from Acoustic BHs}

This breakdown 
of Lorentz symmetry raises the interesting question of whether HR
will be 
modified, given that Hawking phonons emerge
from configurations with wavelength 
reaching the molecular scale.
The outcome of the analysis is that HR is robust \cite{unruh95,bmps}.
That is, to leading order in
the 
parameter 
$\kappa d_c \ll 1$,
the properties of HR, thermality
and stationarity, are preserved. From this we learn the first lesson:
the properties of HR are not tied to the relativistic (scale invariant)
mode equation.

Besides this robustness, there is another important lesson
which is delivered by the analogy. One notices that the
unbounded growth of frequencies  associated with Eq. (\ref{boundlessg})
is exceptional, in that it is found only for the relativistic
dispersion relation $\omega^2 = c^2 p^2$.
For 
{\it all} non-linear dispersion relations,
the new UV scale $p_c= 1/d_c$ 
effectively cuts off the growth of frequencies.
This generic behavior raises doubts about the validity, not on 
the existence of HR, 
but on 
the {\it free} field
dispersion relation $\omega^2 = p^2$, used  to characterize
field propagation in the near horizon region \cite{rpdublin},
see also \cite{sciam} for a broad audience presentation
of this point. 
Indeed the use of this relation amounts to assume
that no UV scale would modify the
two-point correlation 
function
when taking into account
short distance effects.

\subsubsection{Backreaction effects 
on acoustic metrics}

It should be 
emphasized
that HR has been
derived from a {\it linearized} treatment, in which the curved metric entered only as
a passive background. However, in a full 
treatment, the degrees of freedom of the metric 
(or those at the origin of the metric) should also be treated
dynamically and quantum mechanically.
The first step to include the backreaction is clear and rather standard.
It corresponds to the usual secular effects, which occur to second
order in perturbation theory. These secular effects appear classically and quantum
mechanically, both for gravitational BHs and for the acoustic BHs.
It is therefore of value to compare how these
influence both kinds of systems since, \emph{a priori}, the analogy works
 only at the kinematical level.
Indeed,
the evolution of a gravitational black hole is determined by Einstein
equations, whereas the dynamics of a  sonic black hole is governed
by hydrodynamical equations.

\subsubsection{Plan of the paper}

The paper is organized as follows. 
\\
\noindent In section 2 we
describe the classical propagation of sound waves
in inhomogeneous fluid flows.
This provides
the formal structure which governs the hydrodynamical-gravitational analogy.
\\
\noindent In section 3 we recall the basics features of quantum field theory
in a curved metric.
\\
\noindent In section 4 we show how HR
emerges 
when an acoustic BH forms.
\\ \noindent In section 5 the trans-Planckian question is addressed
when taking into account the modification of the dispersion relation at short
distance.
\\ \noindent  Section 6 is  devoted to  the study of
the backreaction on the fluid flow due to the 
energy momentum tensor of the quantum phonon field.
\\
\noindent In section 7 we conclude with some remarks.\\
\noindent
To further appreciate the analogies and the differences between
the gravitational and hydrodynamical cases, we have added
for non-expert readers
an extended Appendix wherein we review the major features of
BHs in
General Relativity.

\clearpage
\section{Sound waves in inhomogeneous flow}

As discussed in the Introduction, Hawking Radiation
is 
not related to the dynamics of
the curved 
metric.
One may therefore
expect to find HR
in different physical contexts whenever
 the propagation of a
 quantum field
 is similar to that in a BH metric.
 The simplest example 
is provided by a fluid whose motion becomes supersonic in some
region.
In a seminal work, dated  1981 \cite{unruh81}, Unruh showed that sound
waves in this system propagate exactly as light does in a black
hole 
metric. For reviews on this topic we refer to \cite{viss02, review, libro}.
In this section we
analyze 
how this result is obtained from a linearized treatment. 
We then study the next order effects
(backreaction).

\subsection{The action principle}

The action principle constitutes the most straightforward way to
derive the equations of motion of a system
(Euler-Lagrange equations)
and the conservation laws associated to its symmetries
(by making use of Noether theorem). This procedure can also be
applied to fluid mechanics. \\  Let us consider a fluid
which is irrotational (i.e. the velocity $\vec v$ can be expressed as
gradient of a potential $\psi$) and homentropic 
(i.e. the pressure
$P$ is a function of the density $\rho$ only). 
In the absence of external
forces, the Eulerian equations of motion for this fluid
can be derived from the action \cite{landau, schakel, stone}
\be \label{hydroaction}
S=-\int d^4 x \left[\rho\dot\psi + \frac{1}{2}\rho (\vec
\nabla\psi)^2 + u(\rho)\right]\ , \ee where $u(\rho)$ is the
internal energy density and a
dot means derivative with respect to newtonian time $t$. \\
\noindent
Variation of $S$ with respect to $\psi$ gives the
continuity equation \be \label{conteq}\dot\rho + \vec \nabla
 \cdot (\rho\vec v)=0 \, , \ee with $\vec v={\vec \nabla}\psi$. Variation of
$S$ with respect to $\rho$ yields Bernoulli equation \be
\label{berneq}\dot\psi + \frac{1}{2}\vec v^2 +\mu(\rho)=0 \, , \ee with
$\mu(\rho)\equiv \frac{du}{d\rho}$.
The gradient of the
Bernoulli equation gives the Euler equations 
\be \label{eulereqs}\dot{\vec v}+ (\vec v \cdot {\vec \nabla} )\vec
v + \frac{1}{\rho}\vec\nabla P =0 \ , \ee where the pressure $P=\int \rho d\mu$.\\
The invariance of the action under translations \bea \label{transl}\psi(x^i) &\to&
\psi(x^i - a^i) \ , \nonumber\\ \rho(x^i) &\to& \rho(x^i - a^i) \ , \eea leads,
through Noether theorem, to the momentum conservation law \be
\partial_t (\rho v_i)+ 
\vec \nabla\cdot ( \vec v
v_i)+\rho
\partial_i\mu =0 \, ,\ee
or, equivalently,
\be\label{Pi_ij}
\partial_t (\rho v_i)+ \partial_j \Pi_{ij}=0 \ , \ee
where $\Pi_{ij}=\rho v_iv_j +\delta_{ij}P$
is the momentum flux tensor and $\rho v_i$ the momentum density.
Finally, the invariance of the action
under rescaling of the velocity potential, $\psi\to \psi +
\alpha$, brings again 
the continuity equation (\ref{conteq}).

\subsection{The 
Acoustic Metric}
\label{unruh}

To 
obtain the propagation of sound waves, 
we expand the action $S$ up to quadratic order 
in the
linear
fluctuations $\rho_1$ and $\
\psi_1$ around some background configuration $\rho_0$ and $ \psi_0$.
This configuration
 describes the mean flow, namely a solution of the
 (unperturbed) 
 equations of motion
(\ref{conteq}, \ref{berneq}).
Therefore, the linear terms in $\rho_1$ and $\psi_1$ drop out,
and one has
 \be S=S_0+S_2\ , \ee where \be S_2= -\int d^4 x \left[ \rho_1 \dot\psi_1 + \frac{c^2}{2\rho_0}\rho_1^2
+ \frac{1}{2}\rho_0(\vec\nabla \psi_1
)^2 + \rho_1\vec v \cdot\vec\nabla
\psi_1 \right]\ \, . \ee
In the latter equation, we have introduced the speed of sound $c$,
which is defined by $c^2=\rho \left. \frac{d\mu}{d\rho}\right|_{\rho_0}$. 
For simplicity, in this paper we shall consider $c=const$.
\\ One can
eliminate $\rho_1$ in favor of $\psi_1$ by using the $\rho_1$
 equation of motion \be
\label{eqrho1}\dot\psi_1 + \vec v_0 \cdot \vec v_1 +
\frac{c^2}{\rho_0}\rho_1 =0 \, , \ee
 and find
\be\label{actpsi1}
S_2 =-\int d^4 x\left[ \frac{1}{2}\rho_0 (\vec\nabla \psi_1)^2
-\frac{\rho_0}{2c^2} (\dot\psi_1 +\vec v \cdot \vec\nabla
\psi_1)^2\right]\, .\ee
This action gives the evolution equation for
 the perturbation of the velocity potential $\psi_1$:
\be
\label{eqpsi1}-\partial_t \left[ \frac{\rho_0}{2c^2}
 (\partial_t\psi_1 + \vec v_0\cdot \vec\nabla\psi_1)\right] +\vec \nabla\cdot
\left\{ \vec v_0 \left[-\frac{\rho_0}{c^2} (\partial_t\psi_1 + \vec
v_0 \cdot \vec\nabla \psi_1)\right] + \rho_0 \vec \nabla\psi_1 \right\} =0\ . \ee
Eqs. (\ref{eqrho1}) and (\ref{eqpsi1})
describe the propagation of linear fluctuations, i.e. sound waves.

The crucial point for us 
is that Eq. (\ref{eqpsi1}) can be  rewritten
in a four dimensional 
notation as \be
\label{eqpsi1f1}
\partial_{\mu}(f^{\mu\nu}\partial_\nu \psi_1)=0 \ ,\ee 
where
\beq\label{fmunu} f^{\mu\nu}\equiv\frac{\rho_0}{c^2}\left(
\begin{array}{cc}
-1& -v_0^i\\
-v_0^i&c^2\delta_{ij}-v_0^iv_0^j
\end{array}\right).\eeq
Moreover, if  we 
set \be f^{\mu\nu}=\sqrt{-g}g^{\mu\nu} \, , \ee then Eq.
(\ref{eqpsi1f1}) 
corresponds to the d'Alembert equation in
the 
curved metric described by $g^{\mu\nu}$ 
\cite{unruh81, visser, wald}, namely
\be\label{waveeqpsi1}
\frac{1}{\sqrt{-g}}\partial_\mu (\sqrt{-g}g^{\mu\nu}\partial_\nu
\psi_1)
 \equiv 
 \Box_g
\psi_1 =0\  \,. \ee
Similarly, the action $S_2$ 
of Eq. (\ref{actpsi1}) can be rewritten as
\be \label{S2} S_2=-\frac{1}{2}\int d^4 x \sqrt{-g}g^{\mu\nu}\partial_\mu\psi_1 \partial_\nu\psi_1 \ .\ee
One can also introduce the covariant 
"acoustic metric"
$g_{\mu\nu}$, which is the inverse of $g^{\mu\nu}$
and reads
\beq\label{gmunu} g_{\mu\nu}\equiv\frac{\rho_0}{c}\left(
\begin{array}{cc}
-(c^2 -v_0^2)& -\vec v_0^T\\
-\vec v_0& \mathbf{1}
\end{array}\right)\ , \eeq
where $\mathbf{1}$ is the identity $3$x$3$ matrix.
The corresponding "acoustic line
element" is
\be\label{acmetric} ds^2\equiv
g_{\mu\nu}dx^\mu dx^\nu=\frac{\rho_0}{c}\left[ -(c^2-\vec
v_0^2)dt^2 - 2 \delta_{ij} v_0^i dx^j 
dt + \delta_{ij}dx^idx^j \right]\ . \ee

In brief, according to Eq. (\ref{waveeqpsi1}),  sound waves
propagation  
coincides with the propagation of a minimally coupled
massless scalar field in the 
curved spacetime described by the acoustic metric $g_{\mu\nu}$.\footnote{The general action for a non-interacting
scalar field $\phi$ propagating in a curved spacetime has the form \cite{bd}
\be S=-\int d^4x \sqrt{-g}\left[\frac{1}{2}g^{\mu\nu}\partial_\mu \phi\partial_\nu
\phi-\frac{m^2}{2}\phi^2+\xi R\right] \, \label{25}\ee where $m$ is the mass of the
field, and $\xi$ measures the non-minimal coupling to the Ricci scalar. 
Eq. (\ref{S2}) corresponds to the minimal coupling massless case, namely $m=\xi=0$.}
 Unlike the particles constituting
the fluid, which see spacetime as 
Galilean and flat,\footnote{Note that in \cite{moncrief}
Moncrief started with a four dimensional relativistic space (flat or curved),
considered a relativistic flow and derived the equations for the
linear perturbations. In this case as well, one finds that
the equations of motion are governed by a four dimensional "acoustic metric".}  sound waves
propagate in a four dimensional curved metric
$g_{\mu\nu}$ (the origin of the curvature
is rooted to the non homogeneity of the flow, see \cite{volovik}).
As we shall see in the following subsection, this
implies that sound propagates along the null geodesics of
$g_{\mu\nu}$ exactly as light moves along the null geodesics of 
spacetime. This 
analogy between propagation of sound
waves and propagation of massless radiation in a gravitational
field allows to translate, using the language of differential
geometry, 
geometrical features of General Relativity into the
 hydrodynamical model. 


It should be also noted that
the three dimensional spatial sections $t=const.$ of the acoustic metric $g_{\mu \nu}$
are conformally flat.
When considering an inward radial flow,
the acoustic metric $g_{\mu \nu}$ resembles to that of a Schwarzschild metric
expressed in the so-called Painlev\'e-G\"ullstrand (PG) coordinates,
for which the spatial sections are also flat
(see Eq. (\ref{pg}) in the Appendix \ref{black.holes}). Note 
that the resemblance is up to an overall 
conformal factor which, as well
known from differential geometry, 
leaves unchanged the equations yielding the null geodesics. Hence, in our case, the conformal factor does not affect the propagation of sound waves \cite{mtw,wald}.

\subsection{Geodesics from geometric acoustics}\label{geod}

In this section, we translate in the acoustic setting the derivation
of the geodesic equations of General Relativity given in \cite{mtw} and in Appendix A.  \\
Let $\lambda$ be the typical wavelength of sound waves and $L$ the
typical length over which the amplitude and the wavelength of the
waves vary.
Geometric acoustic is valid whenever 
$\lambda \ll L$.
\\ \noindent Focussing on waves that are highly monochromatic in a
region $\stackrel{<}{\sim}L$, we split the velocity potential
$\psi_1$
into a rapidly changing real phase $S$ 
and a slowing varying 
amplitude $a$
\be \label{psi1.atheta} \psi_1=\Re \{ a e^{i S}  \} 
\ . \ee
Introducing the wave vector $K_\mu=\partial_\mu S$,
sound rays are 
curves normal to the constant phase
surface $S 
=\mbox{const}$. The differential equations for sound rays
is $dx^\mu/ds =K^\mu$ 
where $s$ is a parameter. Using Eq. (\ref{waveeqpsi1}), one gets
\be \label{nlge} g_{\mu \nu } K^\mu  K^\nu
=0 \, , \ee i.e. $K^\mu$ is a null vector.
After covariant differentiation of the equation above, one can 
 show that $K^\mu$ satisfies the geodesic equation (see Eq. (\ref{KK})
 in Appendix \ref{GR}).
Thus, sound propagates along the null geodesics of the
four dimensional acoustic metric  (\ref{gmunu}).

It is interesting to re-express the above equation in a 3 dimensional language.
Expanding the null condition, and 
using the Newtonian time in  place of $s$, we get
\bea \label{nullgeod} &g_{\mu\nu}K^\mu K^\nu=g_{\mu\nu}\frac{dx^\mu}{dt}\frac{dx^\nu}{dt}
=&\nonumber\\ &=\frac{\rho}{c}\left[-(c^2-\vec v_0^2)-2 \delta_{ij}
v_0^i\frac{dx^j}{dt}+\delta_{ij}
\frac{dx^i}{dt}\frac{dx^j}{dt}\right]=0\ .& \eea
This equation implies that
\be \left| \frac{d\vec x}{dt}-\vec v_0 \right|=c \, ,  \ee
where the norm is taken in Euclidean
space. This is the usual (non-relativistic) equation describing
sound propagating with velocity $c$ with respect to the medium
moving with velocity $\vec v_0$.

\subsection{Pseudo energy-momentum tensor}
The energy-momentum tensor is best defined as the functional derivative
of the relevant action with respect to the background metric
(see for example \cite{land.lif}). In the case at hand, 
two metrics are at our disposal: the real Minkowski spacetime metric
and the acoustic metric defined in Eq. (\ref{gmunu}).
Differentiation with respect to the former leads to the real
energy and momentum (see Eq. (\ref{Pi_ij})), while differentiation
with respect to the acoustic metric leads to the so called {\it pseudo}
energy-momentum tensor (for a detailed discussion see \cite{stone}).

Writing the action $S_2$  in terms of the acoustic metric as in Eq. (\ref{S2}), the pseudo energy-momentum tensor for the fluctuations is
\be \label{pseudo} T_{\mu\nu}=-\frac{2}{\sqrt{-g}}\frac{\delta S}{\delta
g^{\mu\nu}}= \partial_\mu \psi_1\partial_\nu\psi_1
-\frac{1}{2}g_{\mu\nu}\partial^\alpha\psi_1\partial_\alpha\psi_1 \
.\ee 
Its covariant components read
\bea
T_{tt}&=&\dot\psi_1^2-\frac{(c^2-v_0^2)}{2c^2}\left[(\dot\psi_1+
\vec\nabla\psi_1 \cdot \vec v_0)^2-c^2(\vec\nabla
\psi_1)^2\right]\ , \nonumber \\
T_{ti}&=&\dot\psi_1\partial_i\psi_1 -
\frac{v_0^i}{2c^2}\left[(\dot\psi_1+ \vec\nabla\psi_1 \cdot \vec
v_0)^2-c^2(\vec\nabla
\psi_1)^2\right]\ , \nonumber \\
T_{ij}&=& \partial_i\psi_1\partial_j\psi_1
+\frac{1}{2c^2}\delta_{ij}\left[(\dot\psi_1+ \vec\nabla\psi_1 \cdot\vec
v_0)^2-c^2(\vec\nabla \psi_1)^2\right]\, . \label{Tmunu.cov.comp} \eea
Because of the invariance of $S_2$ under diffeomorphisms, this tensor satisfies the     
 "covariant" conservation law
\be
\label{conseqps}
\frac{1}{\sqrt{-g}}\partial_\mu\left(\sqrt{-g}T^{\mu\alpha}\right)+\Gamma^\alpha_{\mu\lambda}T^{\mu\lambda}=0\ ,
\ee where the connection coefficients $\Gamma^\alpha_{\mu\nu}$ are defined as
\be \Gamma^\alpha_{\mu\nu}=\frac{1}{2}g^{\alpha\beta}\left( g_{\mu\beta,\nu}+g_{\nu\beta,\mu}-g_{\mu\nu,\beta}\right)\ .\ee
To appreciate the difference between the true energy-momentum tensor
and $T^{\mu\nu}$, as defined by 
Eqs. (\ref{Tmunu.cov.comp}),
and to understand the meaning of Eq. (\ref{conseqps}), one should note
that for our system there are two distinct operations which can be called "translation" in the (say) $z$-direction:

{\bf a.} 
the fluid, together with any disturbance on it, is translated in the $z$-direction;

{\bf b.}  
the fluid itself is left fixed, while the disturbances 
are translated in the $z$-direction.

The first operation is a symmetry of the action provided the background space is homogeneous. The associated conserved quantity is the true newtonian momentum $\int d^3 x \rho \vec v$. Operation {\bf b} instead is a symmetry only when both the background space and the fluid are homogeneous. In this case the $\Gamma^\alpha_{\mu\nu}$  vanish since the acoustic metric is constant and Eq. (\ref{conseqps}) yields the conservation of pseudo energy-momentum.\\
In the general case, $\Gamma^\alpha_{\mu\nu}\neq 0 $ and Eqs.
(\ref{conseqps}) describe how in an inhomogeneous flow field
energy and momentum are exchanged between the waves and the mean
flow \cite{stone}.


\subsection{Acoustic black holes}\label{abh}

As we have seen in subsection \ref{geod},
sound propagates along the null geodesics of
the acoustic metric $g_{\mu\nu}$ of Eq. (\ref{gmunu}).
The latter shows deep resemblances with the Painlev\'e-G\"ullstrand
form of  
the Schwarzschild
metric (see Appendix \ref{black.holes} Eq. (\ref{pg})).\\
\noindent
For simplicity, we shall not consider radial flows with
spherical symmetry. We consider instead linear flows in the
$z$ direction, and assume no dependence on the
perpendicular directions $x$ and $y$. 
We shall also restrict our attention to the null geodesics which
are parallel to the flow, i.e. $dx = dy =0$.
According to Eq. (\ref{nullgeod}), those obey
\be\label{nulgeoacme}
0=\frac{\rho}{c}\left[-(c^2-v^2)dt^2 - 2vdzdt+dz^2 \right] \ . \ee
When supposing that $v<0$, i.e. the
fluid moves from right to left, the
sound waves propagating downstream follow
\be \frac{dz}{dt}=v-c\, , \ee
whereas the sound waves propagating upstream
obey
\be \label{upwave} \frac{dz}{dt}=v+c \, . \ee

Now, if the fluid motion becomes
supersonic, the surface
$|v|=c$ separates the upstream null geodesics into two classes.
In the region where the motion is
subsonic $|v|<c$, upstream waves can propagate to the right
($\frac{dz}{dt}>0$ in Eq. (\ref{upwave})).
Instead, in the region where the
motion is supersonic, upstream propagating waves are dragged
by the fluid motion to the left ($\frac{dz}{dt}<0$ in Eq.
(\ref{upwave})).
\begin{figure}
\includegraphics[angle=0,width=5.0in,clip]{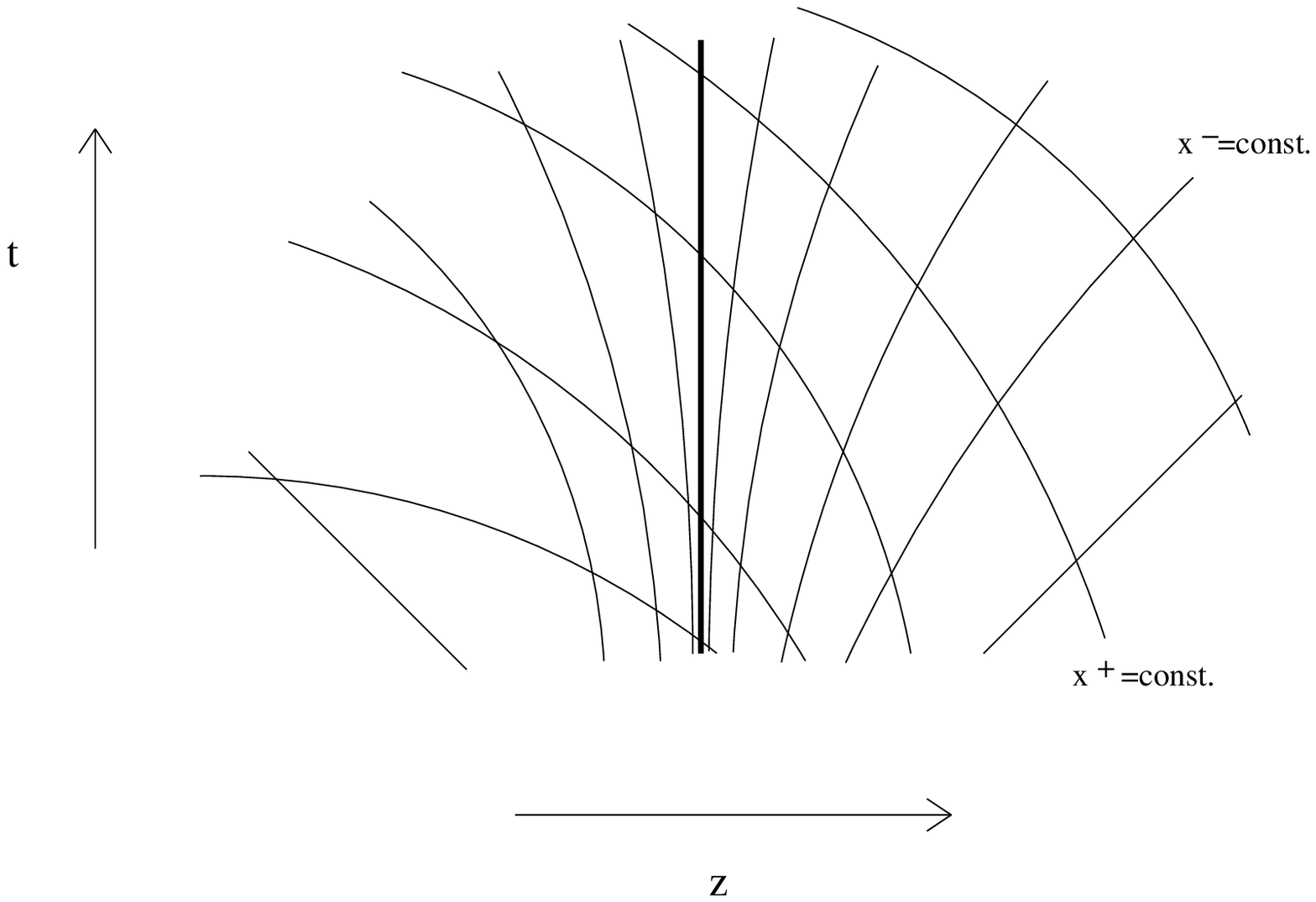}
\caption{
Propagation of sound waves in an acoustic BH.
Families of infalling (downstream) $x^+ = const.$ and outgoing (upstream) $x^- = const.$
null geodesics are depicted.
The vertical
axis is the laboratory time $t$,
and the horizontal one is the $z$ axis.
The fluid
is accelerating and flowing towards the left.
It becomes supersonic at $z=z_h$ which is represented by 
the 
thick vertical line,
the horizon.
On its left
upstream propagating waves are 
swept away by
the supersonic flow and propagate inwards.} 
\label{fig1}
\end{figure}
The surface of the fluid where $|v|=c$
represents the sonic horizon. 
It can be crossed only by downstream propagating sound waves.
Inside this surface, i.e. in 
the supersonic region, sound cannot escape: it is 
trapped by the supersonic motion of the fluid in the same way as light is trapped
inside the horizon of a black hole by the gravitational field.
For these reasons, it is appropriate 
to call this configuration an ``acoustic black hole'' or a ``dumb hole'' \cite{unruh81}
(dumb $=$ unable to speak).
\\ \noindent For stationary flows 
one can introduce null coordinates $x^{\pm}$ \bea x^- &=&
c \left(t-\int
\frac{dz}{c+v}\right)\ , \nonumber \\ x^+ &=&  c\left(
t+\int
\frac{dz}{c-v}\right)\label{xpmnull} \, . \eea
Thus, upstream (outgoing) waves follow 
$x^-=const.$, whereas downstream ones (ingoing)
follow $x^+=const.$\\
A 
  diagram representing the propagation
of these waves is given in Fig. \ref{fig1}.
One should appreciate the strict resemblance 
between this diagram and the
one depicting the behavior of light rays in 
the Schwarzschild black hole geometry
(see Fig. \ref{fig5} in Appendix \ref{black.holes}).
\\ \noindent
It is interesting to further analyze the behavior of the
geodesics in the near horizon region, because it
determines the properties of Hawking 
radiation
when studying quantum effects.
In the near  horizon region 
one can develop the (stationary) velocity profile
as
\be
\label{v(r)}
v(z) = -c + 
c\kappa (z - z_h)  + O(z - z_h)^2 \, ,
\ee
where  $z_h$ is the location of the sonic horizon and we have introduced $\kappa$ which is given by $\kappa = k /c^2$, where $k$ is
the "surface gravity" of the sonic
horizon, defined as \cite{unruh81, visser}
\be \label{surfgra}
k=-\frac{1}{2}\left. \frac{d(c^2-v^2)}{dz}\right|_{z_h} \, . \ee

Inserting the near  horizon expression of $v(z)$ in
Eq. (\ref{upwave}), one sees that outgoing geodesics ($x^- = const.$)
leave the horizon following
an exponential law governed by $\kappa$
\be
\label{explaw}
z(t) - z_h = C e^{c\kappa t} \, .
 \ee
 This implies that the frequency (measured in the rest frame of the
 atoms) of an outgoing sound wave emitted near
the sonic horizon 
will suffer an exponential Doppler shift \be\label{doppshift}
\Omega(t) \sim \Omega_0 e^{- c\kappa t} 
\ , \ee
as it escapes from the hole. (The proof will be given in
subsection \ref{thesm}, 
 see Eq. (\ref{atfreq})).
Comparing these expressions
with those of a radiation field in a gravitational black hole metric
 (Eqs. (\ref{redshiftgrav},
\ref{surgragrav}) in Appendix \ref{black.holes})
one 
verifies that there is a complete
analogy between light propagation in black hole geometry and
sound propagation in an acoustic black hole geometry. 
This should no longer be
a surprise since the field actions and the metrics are identical.

\subsection{De Laval nozzle}
\label{lavalnozzle}
\begin{figure}
\includegraphics[angle=270,width=3.4in,clip]{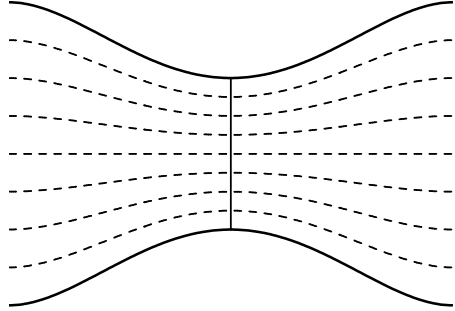}
\caption{A de Laval nozzle.
As explained in the text, it is possible to set the system to have 
a regular flow
that is subsonic on one side of the waist and
supersonic on the other. 
 The acoustic metric associated with
this flow is that of a one-dimensional black hole
whose null geodesics are represented in Figure \ref{fig1}.
The sonic horizon ($|\vec v|=c$) is located at the waist of the nozzle.}
\label{fig2}
\end{figure}

Gravitational black holes are formed by the collapse of very
massive objects ($M\stackrel{>}{\sim} 3M_{sun}$). Sonic black holes
on the other hand do not require such extreme
conditions for their realization. 
They can 
be formed by manipulating laboratory size systems. \\ \noindent
A simple realization is provided by a converging-diverging nozzle, called a 
de Laval
nozzle, well known in jets engine engineering (see for ex. \cite{librohydro}).
It is schematically
depicted in Fig. \ref{fig2}. 
The fluid 
flows from right to left and the nozzle is pointing along the $z$-axis.
Let the cross section area  be $A(z)$. We assume 
that the transverse velocities (in the $x$ and $y$ direction) are
small with respect to the velocity along the nozzle axis (i.e.
$z$). We have then 
a quasi one dimensional flow which we further assume
to be stationary. \\ \noindent In this case the continuity
equation yields \be \label{contnozzle}\rho(z)A(z)v(z)=D\ , \ee
where $D$ is a constant. The Euler equation (\ref{eulereqs}) reads \be
\label{eulernozzle}v\frac{dv}{dz}+\frac{c^2}{\rho}\frac{d\rho}{dz}=0\,
. \ee
By differentiating Eq. (\ref{contnozzle}) to eliminate $\rho$
in Eq. (\ref{eulernozzle}), we obtain \be\label{nozzleeq}
\frac{A'}{A}=\frac{v'}{v}\left( \frac{v^2-c^2}{c^2}\right) \, .\ee
where a prime $'$ means derivative with respect to $z$. 
For subsonic flows (i.e. $|v|<c$) the velocity decreases as the area increases
and vice versa as everyday experience tells us. At the waist of
the nozzle (say $z=0$), $A'=0$ and therefore $v'=0$. 
\\ \noindent
Suppose the fluid starts from $z=+\infty$ with some velocity, say
$v_{+\infty}$. The velocity increases through the converging
nozzle (i.e. $A'/A<0$, $v'/v>0$) reaching the maximum value
$|v_0|<c$ at the waist, and then decreasing to the value
$v_{-\infty}$
along the diverging part of the nozzle ($A'/A>0$, $v'/v<0$). \\
\noindent The picture changes completely when the motion is
adjusted to become supersonic. From Eq. (\ref{nozzleeq}) one sees
that the transition from subsonic to supersonic flow
can occur only at the waist of the nozzle (unless the acceleration
$v v'$ diverges when $|v|=c$, which we consider as unphysical).
Flowing along the divergent part of the nozzle ($z<0$) the fluid
increases its velocity since, for $|v|>c$, $A'$ and $v'$ have
the same sign
in Eq. (\ref{nozzleeq}).  
 This non intuitive
behavior is a well known feature of supersonic fluid motion. The
increase of velocity is associated with a decrease in the density
(and pressure) in the supersonic region ($z<0$).
\\ \noindent The above 
fluid configuration therefore corresponds  to that of a sonic black hole
with
the sonic horizon  located at the waist of the nozzle $z=0$.
The trapped region is the supersonic region ($|v|>c$) on the left
of the waist.
\\ \noindent When the velocity of sound is constant, 
 the nozzle equation (\ref{nozzleeq}) can be immediately integrated
yielding for the acoustic black hole configuration \be
\label{acbhse} A=\frac{c}{|v|}A_H e^{(v^2-c^2)/2c^2}\ , \ee where
$A_H$ is the area of the waist of the nozzle, and
where integration
constants are chosen such that for $A=A_H$ we have $v=-c$.
It is interesting to compare this with the
subsonic solution ($|v|$ always $<c$). In that case,
one has
\be \label{subsonicv}
A=\frac{v_0}{|v|}A_He^{(v^2-v_0^2)/2c^2}\ . \ee
In both cases, 
the
Bernoulli equation gives \be
\frac{v^2}{2}+c^2\ln\frac{\rho}{\rho_0}=0\ , \ee i.e.
\be\label{rhocl} \rho=\rho_0 e^{-v^2/2c^2} \, ,\ee where  the
relation $\mu=c^2\ln \rho/\rho_0$, following from the constancy of
$c$, has been used.
\\ \noindent
The acceleration of the particles of the fluid is given by
$a=vv'$. In terms of this parameter
 the nozzle equation (\ref{nozzleeq}) can be rewritten as \cite{visser}
\be a=\frac{v^2c^2}{c^2-v^2}\frac{A'}{A}\ ,\ee and on the horizon
(using de l'Hopital rule) we have
\be a_H=c^2\left. \sqrt{\frac{A''}{A}}\right|_H \, .\ee
This definition coincides in our case ($c=const.$) with the surface gravity
defined in Eq. (\ref{surfgra}).

We have seen that the
fine tuning condition $A'=0$ forces the acoustic horizon to form
and remain exactly at the waist of the nozzle. The flow will in
fact self-adjust to satisfy this fine tuning which keeps the
acceleration 
finite at the acoustic horizon.
On the other hand, in  presence of an external force $F$ entering Euler
equations (\ref{eulereqs}), the nozzle equation (\ref{nozzleeq})
gets modified, namely \be \frac{A'}{A}=\frac{v'}{v}\left(
\frac{v^2-c^2}{c^2}\right) -\frac{F}{\rho c^2}\ . \ee
The horizon location now satisfies
the equation $A'/A|_H=-F/(\rho c^2)|_H$ and does not coincide
in general with the waist of the nozzle.

\begin{figure}
\includegraphics[angle=0,width=4.5in,clip]{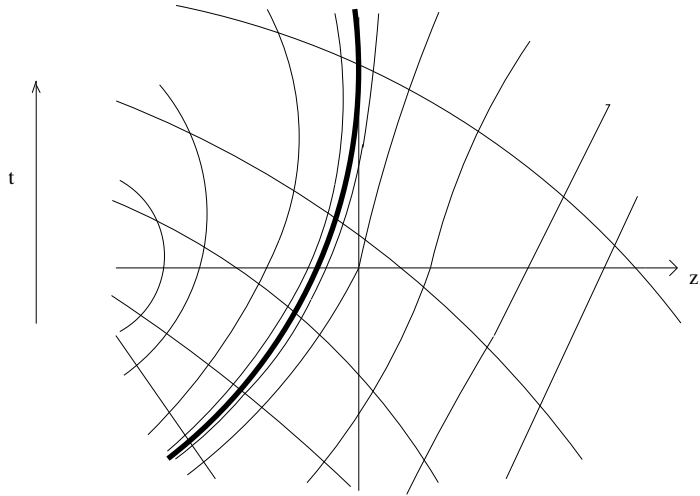}
\caption{Spacetime diagram representing the formation of a sonic black hole.
The solid line denotes the sonic horizon. 
At late times, the acoustic metric coincides with the static one
of Fig. \ref{fig1} as indicated by the trajectories followed
by outgoing null geodesics.}
\label{fig3}
\end{figure}

Finally,
the formation of a sonic black hole can be 
obtained by
a continuous deformation of the velocity profile.
Starting from an initial subsonic configuration, described in
Eq. (\ref{subsonicv}), one obtains the black hole configuration described in
Eq. (\ref{acbhse}) by lowering
the pressure in the exhaust
region (large negative $z$). This evolution is depicted in Fig. \ref{fig3} (see \cite{balisovi}).

\subsection{Backreaction of sound waves}\label{classicalback}

Because of the non-linearity of the
hydrodynamical equations, 
linear perturbations will
backreact on the mean flow.
The dominant effect is
the secular change that they induce.
The backreaction of the waves on the underlying fluid can be
evaluated following the same procedure used to calculate the
backreaction due to the gravitational waves in General Relativity,
as it can be seen from
Appendix \ref{gw} and \cite{mtw}. One writes the physical quantities which describe the fluid as a term
describing the mean flow plus a fluctuation, i.e.
\bea
\psi &=& \psi_B + \psi_1 \ , \nonumber \\ \rho &=& \rho_B + \rho_1 \ . \eea
The subscript $B$ stands for backreaction,
i.e. these quantities include the backreaction
of the waves. The functions $\rho_1$ and $\psi_1$ are
linear in $\epsilon$, where $\epsilon$
is a dimensionless parameter which characterizes the amplitude of the perturbations.
The continuity and Bernoulli equations are then expanded to order
$\epsilon^2$
\bea & &\dot \rho_B + \dot \rho_1 +\vec\nabla \cdot 
\left[ (\rho_B +\rho_1)(\vec v_B
+\vec v_1)\right]=0\ ,\label{backrhob} \\
& & \dot \psi_B + \dot \psi_1 + \frac{1}{2} (\vec v_B + \vec
v_1)^2+\mu(\rho_B)+ \left. \frac{\partial\mu}{\partial\rho}\right|_{\rho_B}
\rho_1 + \frac{1}{2}\left. \frac{\partial^2\mu}{\partial
\rho^2}\right|_{\rho_B} \rho_1^2 =0\ . \label{backpsib} \eea
Since $\r_B$ and $\vec v_B$ contain terms of
order $\epsilon^2$ only, the terms linear in $\epsilon$
must vanish, thus: 
\bea & & \dot \rho_1 +\vec\nabla \cdot
 (\rho_B\vec v_1 + \rho_1\vec v_B)=0\ , \label{backfirst1}\\
& & \dot \psi_1 + \vec v_B \cdot \vec v_1 + \frac{c^2}{\rho_B}\rho_1=0\, .
\label{backfirst2} \eea
As we have seen in subsection \ref{unruh}, these
lead  to the wave equation for a
massless scalar field $\psi_1$.
The novelty 
is that now the d'Alembert operator $\Box_g$ is 
evaluated in the ``mean'' acoustic metric determined by $\rho_B$ and $v_B$.

  From Eq. (\ref{gmunu}), we have \beq\label{gmunub}
g_{\mu\nu}^B
\equiv\frac{\rho_B}{c}\left(
\begin{array}{cc}
-(c^2 -v_B^2)& -\vec v_B^T\\
-\vec v_B& \mathbf{1}
\end{array}\right)\ . \eeq
Taking into account Eqs. (\ref{backfirst1}, \ref{backfirst2}),
the mean flow $(\rho_B, v_B)$ satisfies the backreaction equations
Eqs. (\ref{backrhob}, \ref{backpsib}) without the linear terms in
$\epsilon$:
\bea \dot\rho_B + \vec\nabla \cdot (\rho_B\vec v_B) +
\vec\nabla \cdot (\rho_1\vec v_1) =0\ , \label{backequno1}\\
\dot\psi_B + \frac{1}{2}\vec v_B^2 + \mu(\rho_B) + \frac{1}{2}\vec
v_1^2 - \frac{c^2}{2\rho_B}\rho_1^2 =0 \, .\label{backequno2}\eea 

There is an elegant way to rewrite these 
equations
in terms of the four dimensional 
 pseudo energy-momentum tensor of the fluctuations (see Eq. (\ref{pseudo})),
now evaluated with respect to the mean acoustic metric $g^B_{\mu\nu}$.
We first note that \bea & &\rho_1v_1^i
=-\frac{\rho_B}{c^2}(\dot\psi_1 + \vec v_B \cdot \vec v_1)v_1^i\nonumber
\\ & &=-\frac{\rho_B}{c^2}(T_{ti}+v_B^j
T_{ij})=\sqrt{-g^B}T^0_i \, , \eea and \be \frac{1}{2}v_1^2 -
\frac{c^2}{2\rho_B}\rho_1^2 = - \frac{1}{2}\frac{\rho_B}{c}T\ ,
\ee where $T=g^{\alpha\beta}_B 
T_{\alpha\beta}$ is the trace of the
pseudo energy momentum tensor. 
Then Eqs. (\ref{backequno1}, \ref{backequno2})
become \bea & & \dot\rho_B + \vec\nabla \cdot (\rho_B\vec v_B)
+ 
\partial_{i} (
\sqrt{-g^B} T^{0}_i)=0\ , \label{backeqstmunu1}\\
& & \dot \psi_B + \frac{1}{2}\vec v_1^2 + \mu(\rho_B)
-\frac{1}{2}\frac{\rho_B}{c}  T  =0\ .
\label{backeqstmunu2}\eea
Taking the gradient of Eq. (\ref{backeqstmunu2}) one
obtains the Euler equation which determines the dynamics of the
mean flow
 \be \label{euler.back}\dot{\vec v}_B + \vec v_B \cdot
\vec\nabla\vec v_B +\frac{1}{\rho}
\vec\nabla P_B -\frac{1}{2}\vec\nabla \left( \frac{\rho_B  }{c}
T
\right)=0 \ . \ee
Notice that
if $c\neq const.$ extra terms will appear in Eq. (\ref{backeqstmunu2},  
  \ref{euler.back}). \\
\noindent
There is yet another concept which should be introduced.
One often deals with an ensemble (stochastic
or quantum) of fluctuations rather than with a single
superposition of fluctuations.
Then, to write the backreaction equations for the mean flow,
one should replace $T_{\mu\nu}$ in the above equations by its
expectation value $\langle T_{\mu\nu}\rangle$
in the ensemble under consideration. These expectation values
are functions of
$\rho_B, v_B$ through Eqs. (\ref{backfirst1}, \ref{backfirst2}),
therefore Eqs. (\ref{backeqstmunu1}, \ref{backeqstmunu2})
are indeed the dynamical equations governing the evolution
of the mean fields $\rho_B, v_B$, in 
the Hartree-Fock approximation.

In conclusion, one sees that 
the continuity equation receives a contribution
from the pseudo momentum density,
while the Bernoulli equation gets a contribution
to $\mu$
proportional to the
four dimensional 
trace of the pseudo energy momentum tensor.

\section{Quantum Field Theory in curved spacetime}

As we have seen the propagation of sound waves in a non homogeneous 
fluid flow
corresponds to 
that of a massless scalar field in a curved spacetime described
by the acoustic metric.
Hence quantization of these sound waves
should be discussed within the framework of QFT in curved spacetime (see e.g. \cite{bd}). 
We outline here this procedure.
\\ \noindent
 We start
 by recalling the canonical quantization of a scalar field $\phi$ in
Minkowski spacetime in the Heisenberg representation, where the
metric tensor is $\eta_{\mu \nu}$
in cartesian coordinates. 
The action of $\phi$ is \be S=-\frac{1}{2} \int d^4x \, \eta^{\mu \nu}
\partial_{\mu} \phi
\partial_{\nu} \phi  \, .  \ee
Hence $\phi$  satisfies the massless
Klein--Gordon (KG) equation \be
\label{waveequationM}
\Box  \phi= \eta^{\mu \nu} \partial_{\mu}\partial_{\nu} \phi=0 \ . \ee
To quantize $\phi$, it is appropriate
to separate the positive and negative frequency solutions
with respect to the Minkowski time $t$.
That is, one writes the operator $\phi$ as
\be \label{modeexpansionM}\phi(t, \vec{x})= \sum_{i}[a_i
u_i(t,\vec{x})+ a_i^{\dagger} u_i^{*}(t,\vec{x})]   \, ,  \ee
where  \be
\label{positivefrequencyM}\frac{\partial}{\partial
t}u_j(t,\vec{x})= -i \omega_{j}u_j(t,\vec{x}) \ , \ \ \ \ \omega_{j}>0 \ .
\ee
The positive frequency
solutions $u_i(t,\vec{x})$ and their complex conjugates
form a complete orthonormal basis with respect to the Klein--Gordon
inner
product\index{Klein--
Gordon product}  defined as \be
\label{KGproductM}(f_1,f_2)= -i\int d^3\vec{x} \, (f_1
\partial_{t} f_{2}^*- f_{2}^*\partial_{t}f_1) \ , \ee
where $f_1$ and $f_2$ are solution of the KG equation.
The Fock space is constructed from 
the vacuum state $|0\rangle$
annihilated by the 
destruction  operators $a_i$ \be a_i |0\rangle = 0 \ . \ee
The one-particle states are obtained by acting with the creation
operators $a_i^\dagger$ on this vacuum. Similarly,  the many-particle states
are constructed by acting repeatedly with the
creation operators. \\
Choosing as the orthonormal basis the  plane wave modes \be
u_{\vec{k}} \equiv \frac{1 }{\sqrt{16 \pi^3 w_k}} e^{-iw_kt +
i\vec{k}\cdot\vec{x}} \ , \ee
where $w_k = \vert \vec k \vert$,
the commutation relations for the
creation and annihilation  operators are  \be [a_{\vec{k}},
a_{\vec{k}'}^{\dagger}]= 
\hbar\delta^{3}(\vec{k}- \vec{k}') \ ,
\ee \be[a_{\vec{k}}, a_{\vec{k}'}]=0=[a_{\vec{k}}^{\dagger},
a_{\vec{k}'}^{\dagger}] \ . \ee

This procedure can be extended to curved spacetimes described by a metric tensor $g_{\mu\nu}$.
The action for a minimally coupled massless scalar field propagating
in a curved spacetime 
is given by Eq. (\ref{25}) and the field equation reads $\Box_g \phi=0$, where $\Box_g$ is the 
d'Alembert 
operator in the metric $g_{\mu\nu}$.\\ \noindent In the canonical
quantization procedure, a crucial difference with respect to flat
spacetime is that in curved spacetime there is
no global definition of time and, therefore, 
there is no natural splitting of modes between positive and negative frequency
solutions. Different choices of positive frequency solutions lead,
in general, to different definitions of the vacuum state \cite{bd}.
\\ \noindent
When there is a timelike vector field
$\xi^{\mu}$ leaving invariant the spacetime metric,
one recovers a notion of positive  frequency solutions:
\be
\xi^{\mu}\nabla_{\mu}u_j= -i \omega_{j}u_j \ , \ \ \ \ w_{j}>0 \ . \ee
When the spacetime is not stationary, one looses the 
above criterium to define a single set of positive frequency modes. However,
when the spacetime possesses asymptotic stationary regions in the
past 
and in the future, 
 two sets of positive frequency modes can be defined.
Let us call $u^{in}_{i}$ the orthonormal set of positive frequency modes
in the 
remote past. We can then expand the field $\phi$ as \be
\label{modeexpansionCSTin} \phi= \sum_{i}[a^{in}_i u^{in}_i+ a_i^{in
\dagger} u_i^{in*}]   \ . \ee
Similarly, let $u^{out}_{i}$ denote
the 
set of positive frequency modes 
in the late future. The corresponding expansion for our field is then \be
\label{modeexpansionCSTout} \phi= \sum_{i}[a_i^{out} u^{out}_i+
a_i^{out \dagger} u_i^{out*}]   \ . \ee
Each set of modes defines its own vacuum state: $|in \rangle$ and $|out
\rangle$ respectively. They are defined by \be a^{in}_i |in \rangle = 0 ,\,\, \forall i \  , \ee
and \be
a^{out}_{i} |out \rangle = 0, \,\, \forall i \ . \ee

 Considering our quantum field to be in  the $in$ vacuum state
$|in\rangle$ (i.e. a state with no particles in the past), the
point is to investigate
the 
physical effects induced by the time dependence of the background metric at
late times, when the geometry has settled down again to a
stationary configuration. 
In general, the state $|in\rangle$
will be not perceived as a vacuum state by 
particle detectors situated at late time. These detectors will indeed react
to the presence of $out$ quanta.
To quantify these effects 
it is 
sufficient to compute the 
Bogoliubov transformation
which relates the two sets of modes \cite{bd}.

Since both sets of modes are complete, one can always expand modes
(and operators) of one set in terms of those of the other set: 
\bea
\label{bogolubovtrans} u^{out}_j &=& \sum_{i} (\alpha_{ji}u^{in}_i +
\beta_{ji}u^{in*}_i) \ ,
\nonumber \\
a ^{out}_j &=& \sum_{i} (\alpha_{ji}^* a^{in}_i - \beta_{ji}^*
a^{in\, \dagger}_i)
\, . \eea
These relations are the Bogoliubov
transformations and the matrices $\alpha_{ij}$, $\beta_{ij}$ are
called Bogoliubov coefficients. They are defined by 
the overlaps
 \be
\alpha_{ij}=(u^{out}_i, u^{in}_j) \ , \    \
\beta_{ij}=-(u^{out}_i, u^{in*}_j) \ , \ee
where $(\, ,\, )$ denotes  the Klein--Gordon  product extended to curved spacetime \be
\label{KGproductCST}(f_1,f_2)= -i\int_{\Sigma} d\Sigma^{\mu} (f_1
\partial_{\mu} f_{2}^*- f_{2}^*\partial_{\mu}f_1) \ . \ee
The symbol $\Sigma$ designates a 
Cauchy hypersurface for  the d'Alembert operator.
The differential $d\Sigma^{\mu}= d\Sigma n^{\mu}$, where $d\Sigma$ is the volume
element and $n^{\mu}$ is a future directed unit normal vector to
$\Sigma$.

The non-vanishing character of the
$\beta_{ij}$ coefficients in Eq. (\ref{bogolubovtrans})
means that the positive frequency 
modes $u^{in}_j$ are not positive frequency in the $out$ region.
This 
implies that the vacuum state $|in \rangle$
contains $out$ quanta. Indeed, using Eq. (\ref{bogolubovtrans}),
the mean occupation number of the
$i^{th}$ out quanta, $ N_{i}^{out} \equiv 
a^{out \dagger}_{i}
a^{out}_{i} $, is
\be \label{Nout}
\langle in | N_{i}^{out}| in \rangle = \sum_{j}|\beta_{ij}|^2 \ .
\ee
Hawking radiation is 
an application of this result. 
In that case,
the difference between the $|in\rangle$ and $|out\rangle$ vacua
is a consequence of the formation of the black hole.

\section{Hawking radiation in acoustic black holes}
\label{HRinacBH}

\subsection{Introduction}
Given the 
equivalence between the action governing 
massless radiation propagating in a 
black hole geometry and that of sound waves in an acoustic black
hole geometry, one expects, when sound waves are quantized,
to obtain Hawking radiation 
in the form of a thermal emission of phonons when a sonic hole
forms
\cite{unruh81}.
We shall outline here how this result is obtained,
giving 
a more detailed derivation in the next section.\\
The phonon field
$\psi_1$
is quantized
exactly like the massless scalar field $\phi$ in the former section. 
Its dynamics is governed by the action of Eq. (\ref{S2}) and it obeys the usual
commutation relations.

Let the acoustic metric $g_{\mu\nu}$ be the one associated to the
flow (from right to left) of a fluid in a de Laval nozzle, as
discussed in
section 2.5. 
Suppose the initial configuration ($t\rightarrow -\infty$) corresponds to a
stationary subsonic flow.
In this case, the normal modes of the field are 
asymptotically 
plane waves $e^{-i\Omega x^+/c}$, $e^{-i\Omega x^-/c}$,
where $x^\pm$ are defined in Eqs. (\ref{xpmnull}).
The former modes propagate downstream, from right to left,
and are Doppler shifted to shorter wavelengths 
when the velocity of the flow increases.
The latter propagate upstream, from left to right,
and are Doppler shifted to larger wavelengths in the same situation.
Both modes are positive frequency with respect to the Newtonian time $t$,
and stay positive frequency as long as the flow stays stationary  and subsonic.
\\ \noindent
Expanding the field in this basis,
 one defines the $in$ vacuum $|in\rangle$ which
contains no quanta (i.e. no phonons) at $t=-\infty$.

Now suppose that the velocity profile varies till an acoustic horizon develops,
 and then settles so as to
 engender 
 a stationary black hole geometry of the form described by Eq. (\ref{gmunu}). 
 When the horizon forms, the modes propagating
 upstream separate into two classes, 
 as clearly seen in Figure \ref{fig3}.
 Those on the left of the
 horizon remain trapped in the supersonic region.
 Instead those on the right succeed in escaping the hole
 and suffer a Doppler shift
 which 
 becomes infinite, as one might guess from Eq. (\ref{doppshift}).
 Just before the formation of the horizon, one can show that
 the Doppler shift exponentially  increases.
This means that an initial plane wave of frequency $\Omega$
behaves at late time and far away from the horizon as $e^{-i\Omega X^-}$,
where $X^-$ scales as
\be X^- \propto 
- 
e^{- \kappa x^-}
\, \ee with $k =c^2\kappa$ the surface
 gravity of the sonic horizon. \\
 \noindent These distorted 
 modes are no longer positive frequency with respect to the time $t$.
In other words there is a non-trivial ($\beta \neq 0$) Bogoliubov
transformation between these $in$ modes and the stationary
$out$ modes $\sim e^{- i \omega t}$ 
of positive frequency $\omega$.
When evaluating these Bogoliubov coefficients, one obtains,
following Hawking \cite{hawking1},
that the mean occupation density (i.e. per unit frequency and per unit time)
of outgoing asymptotic quanta 
obeys the Planck distribution:
\be \label{Nout2}
\langle in | N_{\omega}^{out}| in \rangle = {1 \over e^{\hbar 
\omega/\k_B T_H} - 1}\ , \ee 
where
\be \k_B T_H=\frac{\hbar k}{2\pi c
}\ . \ee
An 
estimate for the order of magnitude of this temperature gives (see Visser's
contribution in \cite{libro}) \be T_H=1.2
\times 10^{-6} K mm \left[\frac{c}{1km/sec}\right]
\left[\frac{1}{c}\frac{\partial(c-v)}{\partial z}\right]\ . \ee
For the case of a supersonic flow of water through a $1 mm$ waist
we have $T_H\sim 10^{-6}K$, while for liquid Helium $T_H\sim 10^{-4}K$. In the
case of Bose-Einstein condensates this temperature is $\sim 10\, nK$, which is
not so low if compared with the usual condensate temperature $\sim 100 \, nK$ (see \cite{bvl}).

\subsection{Formal derivation}
\subsubsection{Introduction}
We give a detailed derivation of the Hawking Radiation
in acoustic black holes
in the usual relativistic way \cite{PhysRep95, viss03} 
in a manner which on one hand reveals the
essential ingredients of this radiation,
and on the other hand will show why 
the properties of HR are robust when using a nontrivial dispersion
relation which breaks Lorentz invariance.\\ In
order to keep the discussion as simple as possible,
we shall adopt several simplifications which do not
affect the key features of HR.

As in section \ref{lavalnozzle}, we assume the acoustic black hole
is described by a quasi one-dimensional 
flow of the fluid along the
axis of the nozzle (the $z$ axis). The corresponding acoustic metric is
\be\label{acousticmetric2D} ds^2=\frac{\rho}{c}\left[-c^2dt^2+(dz-vdt)^2+dx^2+dy^2
\right] \ ,\ee where the relevant variables, $v\equiv v_z$ and $\rho$,
depend only on $t$ and $z$.\\ \noindent
In this metric, the phonon field $\psi$ is described by the action,
see Eq. (\ref{S2}),
\be \label{newact} S= 
\frac{1}{2}\int 
d^4x  \rho 
\left\{ \left[ \frac{(\partial_t+v\partial_z)}{c}\psi\right]^2-
(\partial_z \psi)^2-(\partial_x \psi)^2 -(\partial_y \psi)^2 \right\}\ .\ee
HR 
arises from the near horizon propagation of the modes of the
quantum field $\psi$. In this region ($z-z_h\ll 1/\k$) $\rho$ varies smoothly,
so we approximate it 
 as a constant in the sequel.
This approximation 
implies neglecting of the potential barrier
associated to a varying $\rho$ which causes 
an elastic scattering of the modes.
This 
would reduce the asymptotic flux,
since the transmission coefficient across the barrier is smaller than unity,
without affecting the very essence of HR.
From now on, we shall work with a rescaled field $\phi =
\sqrt{\rho}\psi $.

For further simplicity, we shall consider only waves that do not depend
on the transverse variables $x,y$.
(A transverse momentum can be easily included
as it will be conserved and will act as a mass square
in the equation for the longitudinal part
of the modes).
In brief we shall study the solution of the reduced two-dimensional
equation obtained from Eq. (\ref{newact}), namely
\be \label{rel.mode.eqs}
\left[(\partial_t +\partial_z v(t,z))(\partial_t + v(t,z)\partial_z ) -
c^2\partial_z^2 \right]
\phi(t,z) = 0
\, .\ee
In the next subsection we study the stationary case
wherein $v$ does not depend on time. This is in anticipation
to the fact that Hawking Radiation is a stationary process.

\subsubsection{The stationary modes}\label{thesm}
Because of time translational invariance,
it is appropriate to decompose the scalar field as
 \be 
\phi(t,z) =\int _0^{\infty}d\omega \left( \phi_\omega(z)e^{ -i\omega t}
+h.c.\right) \, .\ee
The stationary modes $\phi_\omega(z)$ obey,
see Eq. (\ref{rel.mode.eqs}), 
\be \label{rel.mode.eqs.sim}
(\omega +i\partial_z v)(\omega +iv\partial_z )\phi_\omega=-c^2\partial_z^2\phi_\omega
\ .\ee
The solutions are orthonormalized by the 
Klein-Gordon
inner product
\be \label{innerpr} \langle \phi_\omega|\phi_{\omega'}\rangle=\int dz \left[ \phi^*_\omega(\omega'+iv\partial_z)\phi_{\omega'}- \phi^*_{\omega'}(\omega+iv\partial_z)\phi_{\omega}\right]\ .\ee
The simplified equation (\ref{rel.mode.eqs.sim}) possesses a two dimensional conformal invariance.
This is reflected 
in the orthonormalized 
solutions, which are particularly simple:
\be
 \phi_{\omega, \pm} = {1 \over \sqrt{2\omega 2\pi}}  \exp(  i{\omega} \int dz g_\pm (z))
\, ,
\label{modesolsimp}
\ee
where
\be
g_\pm(z) = { 1 
\over v(z) \pm c } \, . \footnote{To make contact with Eqs. (\ref{xpmnull}),
note that $x^\pm=ct\pm c\int dz g_\mp $. Therefore $e^{-i\omega t} \phi_\mp=
{e^{-i\omega x^\pm/c}}/{\sqrt{2\pi 2\omega}}$. } 
\ee
In anticipation to the situation where we shall use
non-trivial dispersion relations, let us relate $g$
to the Hamilton Jacobi (HJ) action $S$, a solution of $g^{\mu\nu} \partial_\mu S \partial_\nu S =0 $,
see Eq. (\ref{nlge}).
As we shall see classical considerations phrased in the HJ language
are useful to understand what is modified and what is not 
 when
using a  non-trivial dispersion relation.

When considering  solutions at fixed energy $\omega$,
the HJ action obeys the equivalent of Eq. (\ref{rel.mode.eqs.sim}):
\bea
&& (\omega -  v \partial_z S)^2
  =  c^2 
   (\partial_z S)^2
\, .
\label{modeeqsimp}
\eea
Hence the relation between $g$ and $S$ is
\be
S_\pm(z) =  \omega \int^z\!dz \,  g_\pm(z)  \, .
\ee
These two actions
describe outgoing $+$ and infalling $-$ null radial geodesics.
Explicitly these are obtained from $S$ by considering $W = S - \omega t$
and requiring that $W$ be independent of $\omega$.
The geodesics are obtained as the locus
of stationary phase upon building wave packets in $\omega$ (the geometrical approximation).

It is of interest to explicitly consider the
behavior of
stationary modes in the
near horizon region, wherein $v$ can be accurately approximated by
Eq. (\ref{v(r)}).
In the linear regime,
\be
S_+(z) = { \omega \over c\kappa } \ln(z-z_H) \, .
\ee
This logarithmic behavior weighted by the frequency in the unit of the
surface gravity is the key feature characterizing the phase of (outgoing) modes
when approaching a (future) horizon.
On one hand, at the classical level, $\partial_\omega W =0$
gives $t - t_0 = \ln(z-z_H)/ c\kappa$, i.e., the exponential law of Eq. (\ref{expon})
and Eq. (\ref{explaw}).
On the other hand, this logarithm implies that
the asymptotic outgoing modes of Eq. (\ref{modesolsimp}) are singular
on the horizon and only defined for $z > z_H$.

Hence to obtain a complete set of outgoing modes which covers both sides
of the horizon, one should introduce a new set of outgoing modes living
inside the horizon.
We shall call them  trapped modes.
A convenient basis of these modes is provided by
the mirror reflected solutions of the outside solutions:
\be
 \phi^*_{\omega, trap} = \theta(z_H - z)\times {1 \over \sqrt{2\omega 2\pi}}
 \,  \exp( i
 { \omega \over c\kappa } \ln(z_H - z)) 
\, ,
\label{modesolsimpins}
\ee
where $\theta$ is the Heaviside function.
These modes 
have been complex conjugated,
because they have a negative KG norm when $\omega$ is positive.
In the Dirac language they would describe holes,
i.e., in second quantization,
they must be associated with creation operators.
The origin of the sign flip of the KG inner product
can be seen from
Eq. (\ref{innerpr}): 
inside the horizon, the sign of
$p= -i\partial_z$ flips when acting on $\ln\vert z-z_H\vert$.

One should also notice that the logarithmic behavior of modes
implies that the local value of the 
longitudinal momentum $p$ behaves as
\be
\label{peq}
p = \partial_z S_+ =  { \omega \over c\kappa } \frac{1}{z-z_H} = p_0 \, e^{- c\kappa( t- t_0)} \, .
 \ee
The growth of $p$ as $1/( z-z_H)$
coincides with the growth of frequencies of stationary outgoing modes of Eq.
(\ref{modesolsimp})
as measured in the fluid's atoms frame.
Indeed this $z$ dependent frequency is
\be
\label{atfreq}
\Omega(z) \equiv \omega - v(z) \, p = c p = { \omega \over \kappa } \frac{1}{z-z_H}
\, .
\ee
The first equality gives the relationship between the stationary
laboratory frequency and that in the comoving frame.
The second equality follows from the dispersion relation Eq. (\ref{modeeqsimp}).
It is interesting to point out that Eq. (\ref{atfreq})
also appears in the context of gravitational BHs, see Eq. (\ref{boundlessg}).
In that case it gives the relationship between the stationary
asymptotic frequency $\omega$ and the frequency measured by
a free falling observer in the near horizon region.
With these remarks we have completed the list of classical 
properties
governing outgoing propagation in this region.

Unlike what we just obtained for the outgoing modes,
the infalling modes, the $-$ solutions
 of Eq. (\ref{modesolsimp}), 
 are regular
and cover both sides of the 
horizon.

In summary, having a complete set of stationary modes, which cover both sides,
we can decompose the  field operator as
\bea
&&  \phi =    \phi_- +  \phi_+
 \nonumber\\
 &&  \phi_-  = \int_0^\infty \!d\omega\,
 ( a_{\omega, -}\,  \phi_{\omega, -} \, e^{-i \omega t} + h.c.)
 \nonumber\\
 &&  \phi_+ =  \phi_{outg}    +  \phi_{trap}
\nonumber\\
 &&  \phi_{outg} =
 \int_0^\infty \!d\omega\,
 ( a_{\omega, outg} \, \phi_{\omega,+} \, e^{-i \omega t} + h.c.)
\nonumber\\
 &&  \phi_{trap} =\int_0^\infty \!d\omega\,
 ( a^\dagger_{\omega, trap}\,  \phi_{\omega, trap}^* \, e^{-i \omega t} + h.c.)
 \, .
\label{heisfo}
\eea
In the first line we have decomposed the operator into its
infalling $-$ and outgoing part $+$.
In the third line, we have split the $+$ part into the operators which live
outside and inside the horizon respectively.
The three families of destruction operators $a_{\omega, -},\,
a_{\omega, outg},\,  a_{\omega, trap}$
annihilate the $out$ vacuum at late time.
The Hawking flux will be generated by the action of creation
operators $a^\dagger_{\omega, outg}$. 

\subsubsection{ The 
vacuum state resulting from a collapse}

To obtain the Hawking flux, we now turn to the second aspect
of the calculation, namely the identification of the state of the field
which results from the formation of a sonic black hole starting in the late past
with a subsonic fluid configuration when initially no quanta were 
present.
This state is usually called the $in$ vacuum, a denomination we shall use.

To be able to relate this state to states with a definite particle content
built with the $a^\dagger_{\omega}$ operators acting on the $out$  vacuum,
we need to consider a 
metric which describes a black hole formation.
Since we shall be interested only in the late time behavior of Hawking radiation,
we can choose the simplest model to describe it.
Indeed, any other regular way to form the same asymptotic metric 
will differ from this one by transient effects
governing the outset of Hawking radiation.
The regularity of the process 
ensures that these transients fade out exponentially fast in time
in the units of $1/ c\kappa$.

The pedagogical model we shall use
describes the formation of the sonic hole by a sudden transition
between subsonic and supersonic motion. For simplicity we assume
that the transition  occurs at a given time $t_0$.
We assume that initially the fluid is at rest (i.e. $v=0$), but we could have
equally attributed it a finite velocity without affecting the
result.
Then
an instantaneous
transition causes the formation of the sonic hole in a
  de Laval nozzle. So for $t<t_0$
 $v=0$, whereas for  
  $t>t_0$,
  $v(z)$ is given by the stationary sonic black hole solution of Eq.
  (\ref{acbhse}).
\begin{figure}
\includegraphics[angle=0,width=3.8in,clip]{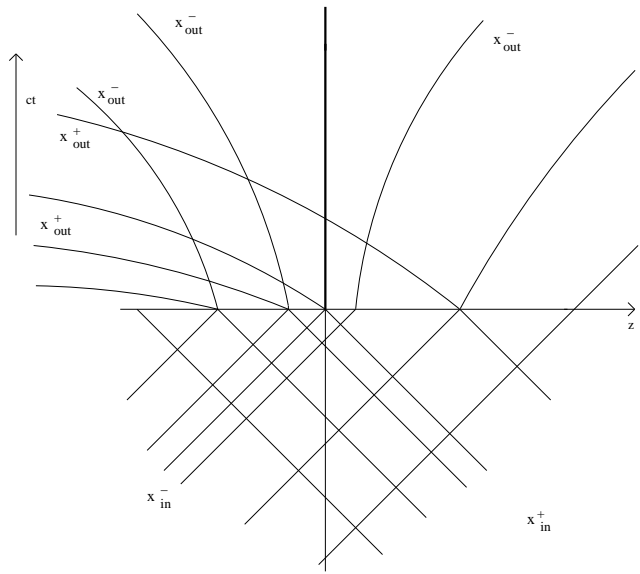}
\caption{Spacetime diagram of an acoustic black hole
produced by an instantaneous shock at $t=t_0$.
Before the shock the null geodesics propagate like
those in flat Minkowskian spacetime, 
but after 
they are no longer described as straight lines since the spacetime
has been modified by the shock and is described by the curved
acoustic metric (\ref{gmunu}) (the vertical solid line represents
the sonic horizon). At late times, the outgoing null geodesics
coincide with those of Figures \ref{fig1} and \ref{fig3}. }
\label{shock}
\end{figure}
A spacetime diagram of the model is shown in Fig. \ref{shock}.

The acoustic spacetime is flat for $t<t_0$.
Hence vacuum positive frequency outgoing modes are simply
plane waves \be \label{Mink.plane.waves}\phi_\Omega=\frac{1}
{\sqrt{2\pi 2\Omega}}e^{-i\Omega t}
e^{i\Omega z/c} 
=\frac{1}{\sqrt{2\pi 2\Omega}}e^{-i\Omega \, x^-_{in}/c} \, .\ee
To find Hawking radiation
one needs simply to paste these modes to the stationary outgoing modes
$\phi_{\omega, +}, \, \phi_{\omega,  trap}^\star$ 
of Eqs. (\ref{heisfo}) which live on either side of the horizon.
This is done in two steps.
First by constructing the $in$ modes
after  the shock
and then by computing
the overlap of the latter with the stationary modes. 

The result is the following. (The proof is given afterwards.)
In the future of the shock, unlike the outgoing modes with fixed
frequency $\omega$, 
$in$
modes 
are regular across the horizon $z=z_h$. 
More precisely, at fixed $t$ they 
are still of the form, see Eq. (\ref{Mink.plane.waves}),
\be
\label{inmodes}
\phi_{\Omega} \simeq e^{i\Omega z/\gamma} \, ,
\ee
where $\gamma > 0$ is a constant,  up to non-universal correction
terms which vanish when $z\rightarrow z_h$. These describe
the outset of HR in the particular formation model one is considering.

 From Eq. (\ref{inmodes}), we learn that
$in$ modes contain only positive momentum $p=-i\partial_z$ in the
near horizon region. This extreme simplicity allows to {\it
characterize} the $in$ vacuum in several instructive ways.
Moreover these only refer to the near horizon stationary behavior
of $in$ modes and thus no longer refer to the region before the
shock.
Nevertheless
to consider the latter
was necessary to {\it derive} the regular properties of $in$ modes.
At this point it is interesting to mention that this will no longer
be the case when considering non-linear dispersion relations.
In these cases, it is possible to derive HR
without any reference to the formation of the BH.

[{\it{Proof of Eq. (\ref{inmodes}).}}
For the first step one needs the relation between laboratory
$t,z$ coordinates and the null coordinates $x^-$ in both regions.
We have, for $t < t_0$:
\bea \label{x-in} x^-_{in}&=&ct-z\eea
whereas for $t>t_0$, in the near horizon region we have:
\bea \label{x-out} x^-_{out}&=&ct-c\int \frac{dz}{c+v}
\simeq ct-\frac{1}{\k}\ln \left(z-z_h\right) +O(z-z_h)\, .
\eea
Evaluated
along the shock these become
\bea x^-_{in}&=& ct_0- z \, , \nonumber \\
x^-_{out}&\simeq& ct_0-\frac{1}{\k}\ln (z-z_h) +O(z-z_h)\ ,
\nonumber \\
&=& -\frac{1}{\k} \ln (x_-^H-x^-_{in})+const \ ,\label{x-out2}
\eea
where $x_-^H= ct_0- z_h $  is the value of $x^-_{in}$
at which the horizon forms. In the last equality
 we have used Eq. (\ref{x-in}) to replace $z$ by $x_{in}^-$.
\\
Eq. (\ref{x-out2}) gives the desired relation $x^-_{out}(x_{in}^-)$ which
allows to extend the null geodesics $x_{in}^- = const.$ in the future of the shock.
One has
\be x_-^H- x_{in}^- \propto e^{-\k x_{out}^-} \propto (z-z_h) \, e^{-c\k t} \, ,
\ee up to irrelevant constants.
This relation holds in the near horizon limit whatever (regular) 
model for the sonic hole
formation is used. At fixed $t$, this implies that an initial mode
$\sim e^{- i \Omega x_{in}^-/c}$ behaves as specified in Eq.  (\ref{inmodes}).
 QED.]

\subsubsection{Near horizon 
$in$ vacuum and HR}

To efficiently characterize the $in$ vacuum and derive
the properties of HR,
it is particularly appropriate to work in the $p$-representation,
even though a priori this might  seem rather abstract. 
In fact, these settings are also appropriate to
re-derive HR 
using a non-trivial dispersion relation, and to
understand the origin of its robustness \cite{bmps}.


The Fourier transform of a mode
\be
\varphi_\omega(p) = \int dz \frac{1}{\sqrt{2 \pi}} \, e^{-ipz} \, \phi_\omega(z) \,
\ee
obeys
\be\label{eq.motion.fourier}
(\omega - p \hat v  )(\omega -  \hat v p) \varphi_\omega =  c^2 p^2 \varphi_\omega\, ,
\ee
where $\hat v = v(\hat z)$ is a differential operator acting on its right.
In the near horizon region, for $z_h=0$,
one has $\hat v = -c + i c
 \kappa \partial_p$ (see Eq. (\ref{v(r)})).
Thus 
Eq. (\ref{eq.motion.fourier}) is a
second order differential equation in $\partial_p$.
In this region, the general solution describing outgoing modes is thus
\be
\varphi_\omega(p) = A_\omega \, \theta(p) \, p^{- i \omega/c\kappa - 1}
+ B_\omega \, \theta(-p)\, ( -p)^{- i \omega/c\kappa - 1}
\, .
\ee
Using Eq. (\ref{innerpr}), one finds that
the Klein-Gordon norm of this solution is
\bea
\scal{\varphi_{\omega}}{\varphi_{\omega'}} = \int dp \, {2}{p} \,  {\varphi^*_{\omega}}
 \varphi_{\omega'} = (|A_\omega|^2 - |B_\omega|^2 )\, (2 \pi 2 c\kappa )
 \, \delta(\omega- \omega') \, .
\label{KGnorm}
\eea From this norm,
we learn that  positive norm (outgoing) modes are principally made with
positive momentum.

The $in$ modes are the simplest ones: they are made exclusively of
positive momentum, i.e., $B_\omega =0$ for both signs of $\omega$,
in agreement to what was find in Eq. (\ref{inmodes}). 
The normalized
$in$ modes of eigen-frequency $\omega$ are thus
\be
\varphi_{\omega, in}(p) = \theta(p)\,
\frac{1}{\sqrt{2\pi 2c\kappa}} p^{- i \omega/c\kappa - 1} \, .
\ee

It is now straightforward to get the coefficients of the Bogoliubov transformation
relating these modes to the singular asymptotic modes.
It suffices to inverse Fourier transform 
$\varphi_{\omega, in}(p)$.
Let us proceed in two steps:
\bea
\phi_{\omega, in}(z) &=& \int_0^\infty dp \, \frac{1}{\sqrt{2\pi}} \, e^{ipz} \,
\frac{1}{\sqrt{2\pi 2c\kappa}} p^{- i \omega/c\kappa - 1} \,
\nonumber\\
&=&
\frac{1}{2 \pi \sqrt{ 2c\kappa}} \, \Gamma(\frac{ - i \omega}{c\kappa})\, e^{\pi \omega/2 c\kappa}
(z+ i\epsilon) ^{i \omega/c\kappa } \, .
\label{inv.fourier}
\eea
In the second line, $\epsilon$ is a positive infinitesimal number.
This equation tells
us that the $in$ modes are analytic in the upper half $z$ plane, a condition which
directly follows from the fact that only positive $p$ entered in the Fourier
transform.
This is crucial:
it tells us how to connect the singular outside and trapped 
modes of Eqs. (\ref{heisfo})
by specifying that the cut of 
ln$z$ is such that the argument
of negative $z$ is $+ i \pi$. This fixes the relative weight of the
 modes living on either side:
\bea
\phi_{\omega, in}(z) &=& \frac{1}{2 \pi \sqrt{ 2c\kappa}} \,
 \Gamma(\frac{-i \omega}{c\kappa})\, e^{\pi \omega/2c \kappa}
 \left[ \theta(z) z^{i \omega/c\kappa } + e^{-\pi \omega/c\kappa}
 \theta(-z) (-z)^{i \omega/c\kappa }
 \right] \, ,
 \nonumber\\
 &=&
\left[ \frac{1}{ \sqrt{2\pi c\kappa/\omega }}
\Gamma(\frac{-i \omega}{c\kappa})\, \right]
\nonumber\\
&& \quad \times \left[ e^{\pi \omega/2 c\kappa} \, \theta(z)  \phi_{\omega,  +}
+  e^{-\pi \omega/2c\kappa}
\,  \theta(-z) \phi^*_{\omega,  trap}
\right] \, , \nonumber\\
 &=&  \alpha_\omega \, \phi_{\omega,  +}  +   \beta_\omega \, \phi^*_{\omega, trap}
 \, .
\label{phi.in(z)}
\eea
In the second line we have used the normalized outgoing modes of
Eqs. (\ref{modesolsimp}, \ref{modesolsimpins}).
The Bogoliubov coefficients are, by definition, the coefficients of these modes.
They obey
\be
| \beta_\omega/ \alpha_\omega|^2 = e^{- 2\pi \omega/c\kappa}\, .
\label{betal}
\ee
Together with the unitary relation $\vert \alpha_\omega \vert^2 -
\vert \beta_\omega \vert^2 = 1 $, one
gets Hawking's result:  the emission rate
of outgoing quanta per unit asymptotic time and per unit frequency
is given by\footnote{
For the unfamiliar reader, we note that, for free fields,
the relationship between $in$ vacuum
expectation values of observables and Bogoliubov coefficients is universal.
Of central interest in stationary cases
is  
 the mean particle density flux $\bar n_\omega$,
i.e. per unit time and per unit frequency, in the $in$ vacuum.
 It is
 given by the weight of the delta in $\bra{0, in} a^\dagger_\omega a_{\omega'}\ket{0, in}
 = \bar n_\omega \delta(\omega -\omega')$.
 In terms of the Bogoliubov coefficients, one has
  $\bar n_\omega =  \vert \beta_\omega \vert^2$. For a detailed derivation of
  these relations,
see \cite{PhysRep95}.}
\be
\bar n_\omega = \vert \beta_\omega \vert^2 = (e^{2\pi \omega/c\kappa} -1 )^{-1}\, ,
\ee
that is, a Planck distribution with a temperature given by $T_H = \hbar c\kappa / 2\pi\k_B$.

There is an interesting way to re-write Eq. (\ref{phi.in(z)})
which is useful to understand the robustness of HR.
It concerns the decomposition of asymptotic modes
in terms of $in$ modes. The very fact that  the asymptotic
stationary modes vanish inside the horizon, for $z< 0$,
implies that they are analytic in the lower half $p$ plane.
Indeed one has
\bea
\varphi_{\omega, \, +}(p) &=&
\int^\infty_0 dz \, e^{ -ipz} \phi_{\omega, \, +}(z) \, , \label{truerel} \\
 &=& \frac{1}{2 \pi \sqrt{ 2\omega}} \,
 \Gamma(\frac{i \omega}{c\kappa}+ 1)\, e^{\pi \omega/2 c\kappa}
 \left[ (p + i \epsilon)^{-i \omega/c\kappa -1 } \right] \, ,
 \nonumber \\
 &=& \frac{1}{2 \pi \sqrt{ 2\omega}} \,
 \Gamma(\frac{i \omega}{c\kappa}+ 1)\, e^{\pi \omega/2 c\kappa}
 \nonumber \\
 && \quad \times \left[ \theta(p) p^{-i \omega/c\kappa -1 } -
 e^{-\pi \omega/c\kappa}  \theta(-p) (-p)^{-i \omega/c\kappa -1}
 \right] \, \nonumber .
\eea From
this equation one gets the Bogoliubov coefficients of the inverse transformation.
They also obey Eq. (\ref{betal}).
Hence HR can be equally deduced from the above relation.
In fact Hawking originally derived black hole
radiance in a similar way \cite{hawking1}.
He propagated backwards in time asymptotic outgoing modes
and decomposed them in terms of 
 $in$ modes to get the Bogoliubov coefficients.


We have now all the tools to address the calculation
of HR with the relativistic equation replaced
by a non-linear Dispersion Relation (DR in the following).
Before that, let us list the essential properties of modes
which fix the properties of HR, namely thermality and stationarity.

{\bf A.} The modes characterizing $in$ vacuum are regular on the horizon and contain
only positive momentum $p$ at fixed time,
see Eqs. (\ref{inmodes}, \ref{inv.fourier}).

{\bf B.} The stationary modes associated with outgoing asymptotic quanta
are singular on the horizon and vanish inside it.
The singularity of a mode of frequency  $\omega$ is governed by a logarithm of
$z-z_h$ weighted
by $\omega/c\kappa$, with $k=c^2\kappa$ the surface gravity.
Since these modes vanish inside the horizon,
in the $p$-representation, they are analytic in the lower
$p$-plane,
see second line of Eq. (\ref{truerel}).

{\bf C.} This analytical behavior fixes the ratio of
Bogoliubov coefficients relating the asymptotic
modes to the regular modes characterizing the $in$ vacuum, see
third line of Eq. (\ref{truerel}) and Eq. (\ref{betal}).\\

These three properties will be essentially {\bf unmodified}
when replacing the relativistic equation (\ref{rel.mode.eqs}), by that
associated with a non-linear DR.
This
explains why the properties of Hawking radiation
are robust against introducing a non-linear DR in the UV sector,
i.e. for frequencies much higher than $c\kappa$.

The usefulness of the $p$-representation
can now be appreciated.
In the near horizon region,
the mode equation (\ref{eq.motion.fourier}) in Fourier transform is
second order, as in the $z$-representation.
However, the symmetry between the $z$ and $p$ representations
is broken upon introducing a non-linear DR:
in the $z$-representation the order of the differential equation will
depend on the choice of the DR, whereas in the $p$ representation,
the differential equation will stay second order.
This allows to perform a general
WKB analysis to interpret the corrections and to evaluate them.
Secondly, for all DRs, the condition 
 for being in the $in$ vacuum
is still given by imposing that the modes 
 only contain positive momentum.
\\ The important consequence of these two facts,
is that the {\it only} source of possible modifications of HR
arising from the near horizon region is backscattering in $p$-space.

\section{Short distance effects and dispersion}
\label{modmodeq}

The first idea is to replace the relativistic relation, which reads
\be
\Omega^2 = c^2 p^2 = c^2 
\, \vec p \cdot \vec p \, ,
\ee
in flat spacetime, by a non-linear
dispersion relation
\be\label{OmegaF}
\Omega^2 = c^2 F^2(p^2)
\, ,
\ee
which breaks Lorentz invariance, but which preserves the
other symmetries of Special Relativity, including
 isotropy.
The non-linear behavior of $F$ introduces a UV scale $p_c$ which weights
the first non-linear term. Generically one has only two cases
\be\label{F}
 F^2(p^2) = p^2 \pm {p^4 \over p_c^2} + O(p^6 )
\, .
\ee
The $+$ case gives rise to superluminous propagation (faster
than light or sound), whereas the $-$ case leads to sub-luminous
propagation.
This follows
from the relation giving the group velocity: $v_g (p)= c 
dF/dp$.\footnote{There
is a also third situation, which corresponds to dissipation.
This case has not been fully analyzed even though it has
been represented in \cite{bmps}
and schematically described in \cite{BarraFrolParent2}.
The main difference between dispersion and dissipation
is that the dissipative case is not autonomous: the environment giving rise to
the dissipation should be explicitely included in order to get a unitary description
of this case. We hope to describe this case in the near future.}

The introduction  of the non-linear function $F$
can be conceived from two different point of views.
First, from the acoustic point of view,
when considering the propagation of collective excitations in condensed matter,
 there is dispersion at short wavelength. 
 In other words, the Eulerian action of Eq. (\ref{hydroaction}) governs
 only the evolution of long wavelengths configurations.
For example, superluminous dispersion is found in dilute Bose gas condensate
\cite{stringaritrentormp},
whereas the subluminous case is found in $\,^4$He, see \cite{ted91,volovik}.
The second point of view is more conjectural and
related to the short distance behavior of
space-time itself. Certain authors consider that Lorentz invariance
might only be an approximate symmetry in nature \cite{ted91,jacobson2},
 and will therefore be violated in the UV, say near the Planck scale.

The second idea is to introduce this modified dispersion relation in a BH spacetime.
Since the modified DR breaks Lorentz invariance locally, one must choose a
particular frame. The choice of this frame 
differs according to the point of view one adopts.
When one considers sound propagation,
there is no ambiguity:
the preferred frame is that
of the atoms of the fluid \cite{ted91,unruh95}. In other words
$\Omega$ in Eq. (\ref{OmegaF}) is the frequency measured
in the frame of the fluid's atoms.
The alternative point related to Quantum Gravity
is again more conjectural.
It is based on the double assumption that the Quantum Gravitational effects,
which have been so far neglected, will preserve the regularity of the $in$ vacuum
across the horizon, but will also
effectively introduce a non-linear DR. In this second point of view,
one should choose what is the preferred frame, or better, the preferred class of frames,
since two frames in a same class will give essentially the same
results. The choice usually made is that the prefered frame
is a freely falling frame. This amounts to introduce the modified DR in the PG
coordinate systems, see Eq. (\ref{pg}) in Appendix \ref{black.holes}.

No matter which  point of view is chosen,
one deals with the following action.
Under the same simplifying hypothesis of the previous section,
the modified action related to the DR of Eq. (\ref{OmegaF}) is
\bea
S &= & \frac{\rho}{2 c^2}  \int dt dz [(( \partial_t + v \partial_z ) \psi )^2
- c^2 \psi F ^2(- \partial_z^2)\psi ]
\, .
\eea
Since the modification concerns only the spatial part of the action,
the inner product and the normalization condition, Eq. (\ref{innerpr}), are
both left unchanged.

The modified mode equation, replacing Eq. (\ref{rel.mode.eqs.sim}), becomes
\be
(\omega + i \partial_z v)(\omega + i v \partial_z) \phi_{\omega} =c^2 F^2( - \partial_z^2) \phi_{\omega}
\, .
\label{modeeqsimpmod}
\ee
To study the near horizon behavior of mode propagation,
we return to the $p$-representation.
The modified equation in Fourier transform reads
\be\label{modif.Fourier.eq}
(\omega - p \hat v  )(\omega -  \hat v p) \varphi_{\omega, F}=  c^2 F^2(p^2) \,
\varphi_{\omega, F}
\ee
where $\hat v = -c + i  c
\kappa \partial_p$, as in Eq. (\ref{eq.motion.fourier}).

The structure of this equation guarantees that for all $F$
the solutions can be written as a product
\be
 \varphi_{\omega, F} =   \varphi_\omega \times e^{-ip/\kappa}\, \chi_F\, ,
\label{factoriz}
\ee
wherein the first factor is the solution of the relativistic equation, hence $F$ independent,
whereas the second factor is $\omega$ independent. 
The function $\chi_F$ obeys the simple second order differential
equation:
\be
\kappa^2
p^2\,  \partial_p^2\,  \chi_F ={ F^2 } \, \chi_F \, .
\label{chieq}
\ee
Moreover, when $\varphi_{\omega, F}$ is conventionally normalized
with the KG product (which is $F$ independent), see Eq. (\ref{KGnorm}),
$\chi_F$ must have the following unit Wronskian
\be
- \frac{\kappa}{2} (\chi_F^* \, i\partial_p \chi_F -
\chi_F \, i\partial_p \chi_F^*) = \pm 1 \, .
\label{wronchi}
\ee
The $+$ sign applies to outgoing modes, whereas the $-$ sign applies to infalling modes.

The equations (\ref{chieq}) and (\ref{wronchi})
fully govern the modifications of HR  which
come from the near horizon region.
Of course there will be other corrections
which arise from the outer region wherein the gradient of the
velocity goes down to zero, see e.g. \cite{cj}. However these modifications
will affect the propagation of modes irrespectively of HR,
as the `grey body factor' does it in the Schwarzschild metric in the
3-dimensional case. In what follows we shall only discuss the
(intrinsic) modifications of HR
which arise from the near horizon region.

To present the nature of these modifications,
we shall split the presentation into three parts.
First we discuss the mathematical properties of $\chi_F$
and its WKB approximation. 
Then we present the calculation
and the interpretation of the corrections to this approximation.
At the end only we argue why the asymptotic conditions to
determine the
modes which describe $in$ vacuum and asymptotic quanta
are both unaffected by the non-linear behavior of $F$.

\subsection{ Mathematical properties and WKB approximation}

Let  us consider the exact solution of Eq. (\ref{chieq})
which for $p \to \infty$ is purely `forward',
i.e. it is a running wave which possesses a unit
positive Wronskian of Eq. (\ref{wronchi}).
When evaluated for moderate values of $p$,
i.e. $p/p_c \ll 1$, the exact solution will be of the form
\be
\chi_F = A_F \chi_{F}^{WKB} + B_F \chi_{F}^{WKB,\, *}  \, ,
\label{wronchi2}
\ee
where $\chi_{F}^{WKB}$ is the following unit Wronskian WKB solution
\be
\chi_{F}^{WKB} = \frac{1}{\sqrt{F(p)/p}}\, exp\left(i \int^p \frac{dp' F(p'^2)}{\kappa p'}\right) \, .
\label{wronchi3}
\ee
The conservation of the Wronskian guarantees that
\be
| A_F|^2  - |B_F|^2 = 1 \, .
\label{wronchi.cons}
\ee

In the WKB approximation, $B_F = 0$: no backscattered wave (in $p$) is engendered
in the journey from $p = \infty$ down to $p \ll p_c$.
In this regime, 
the norm of $A_F$ stays
constant: $\vert A_F(p) \vert = \vert A_F(p=\infty) \vert = 1$.
Hence, there will be no modification of HR in this WKB regime.
Indeed the ratio of the Bogoliubov coefficients, determining Hawking temperature,
is extracted from the {\it relative weight} of the positive and negative
norm modes and not from the local behavior of the modes.
This local behavior is in fact affected by
the non-linear behavior of $F$: up to a constant phase one has
\be
e^{-i p/\kappa}\, \chi_{F}^{WKB} = 1 + O(p/p_c)^2 \, .
\ee
The important point is that
the correction terms in $(p/p_c)^2$ should not affect
the Bogoliubov coefficients of Eq. (\ref{phi.in(z)}).

\subsection{ Backscattering and modifications of HR}

The 
modifications to HR come from
the {\it global} corrections to the WKB approximation,
i.e. the amplitude $B_F$. Indeed, there are two kinds
of modifications to $\chi_{F}^{WKB}$: first there are
local (in $p$) 
modifications which follow from the fact that
$\chi_{F}^{WKB}$ does not exactly obey Eq. (\ref{chieq}) at that $p$.
These corrections should not affect HR,
see the above discussion.
Secondly 
there are global (accumulated) errors which result from the fact that
$\chi_{F}^{WKB}$ did not exactly obey Eq. (\ref{chieq})
from $p= \infty$ down to values $p\simeq \kappa$.
This second type is responsable for $B_F \neq 0$
in Eq. (\ref{wronchi2}).

To evaluate $B_F$ 
is not an easy task
because it is an accumulated error,
see e.g. \cite{NP}.
What can be said is that the local source of $B_F$ is \be
\sigma(p) =\k
 \frac{\partial_p (F/p)}{(F/p)^2} \, .
\ee
 This dimensionless quantity is smaller than  $\kappa/ p_c$.
 From this we learn that scale
separation, i.e. $\kappa/ p_c \ll 1$, is, in full generality, a sufficient condition
for having small corrections to HR.

Moreover, since the final value of $B_F$ (i.e. for $p$
of the order of $\kappa$) is expressed as an
oscillatory integral of $\sigma(p)$,
 the integral will generically be smaller than the
integrand ($=\sigma$). To 
illustrate this we 
consider an exactly solvable case:
the quartic dispersion $F^2 = p^2 + p^4/p_c^2$.
In this case, Eq. (\ref{chieq}) describes an
up-side down harmonic oscillator.
Then using the equations of Sections 19.17 to 19.22
of \cite{Abramo} one gets
\be
B_F = 
e^{-\pi  p_c/\kappa} \, ,
\label{wronchi4}
\ee
which is indeed much smaller than $\sigma \simeq \kappa/ p_c$.

The 
lesson is that the (intrinsic) corrections to HR
originate from the backscattered amplitude and not from the
 local corrections to $\chi_{F}^{WKB}$.
The backscattered amplitude is small whenever the UV scale
governing dispersion is much larger than the surface gravity.

A more subtle consequence of Eq. (\ref{wronchi2}) is that $A_F$, the
amplitude of the forward wave, has {\it increased}
when $B_F \neq 0$.
Hence the corrections to HR are related to
another pair creation process. 
The new  Bogoliubov transformation mixes
outgoing and infalling modes, unlike what was found in Eqs. (\ref{phi.in(z)}).
To our knowledge this has not been
described
in the literature.
We hope to report on this in the near future.

\subsection{ The identification of $in$ and $out$ modes}

It remains to verify that the non-linear behavior of $F$
does not affect
the relationship
between the asymptotic physical states and the mathematical
properties of the modes, solutions of Eq. (\ref{modeeqsimpmod}).

Let us start with $in$ vacuum. We claim that the initial
characterization of the  $in$ vacuum is still that the
$in$ modes only contain positive $p$. Let us explain why
in the light of the properties of the classical propagation
in the near horizon region.

The factorization of the modes in the $p$-representation
is recovered in the
HJ formalism in the following way. The modified Legendre transformed HJ action,
$W_F(p)= S_F(x) - px$,  obeys the equivalent of Eq. (\ref{modeeqsimpmod}),
\be
\omega + p\,  (c + c\kappa \partial_p W_F ) = c\,  F(p^2)\,
.
\ee
Its solution is a sum of the relativistic action plus a modified term
independent of $\omega$
\bea
 W_F (p)
  = { \omega \over c\kappa }\ln p +  \int^p dp' {F(p'^2) -p' \over \kappa p'}\, .
\eea
This tells us that, irrespectively of the choice of $F$,
the trajectory in $p$-space is, in the near horizon, the same as
that of the relativistic case, namely Eq. (\ref{peq}).
So, even though the trajectories in spacetime do depend on $F$,
in $p$ space they always obey the exponential $\kappa$ decay. 

It is important to analyze the propagation in  $z$ space to see that
the exponential blue-shift (backwards in time) is cut off by $p_c$,
the UV scale brought in by $F$. 
Eq. (\ref{peq}) translates in $z$ space in the following manner:
the {\it difference} in $z$
between the 
classical trajectories with positive and negative 
energy $\omega$ always obeys
\be\label{Delta(z)}
z_{F, \omega}(t) - z_{F, -\omega}(t)= \Delta z_{F, \omega}(t) = \Delta z_{F=p, \omega}(t) = {2  \omega \over c\kappa  p_0} \, e^{ c\kappa t} \, .
\ee
Therefore the characteristics of the modified dispersion relation are
still turned apart by the surface gravity,
as they were in the relativistic case. However,
the 
important
 difference is that this stretching
lasts only a finite 
time for non-linear dispersion, 
either super- or sub-luminous.
To prove this we consider the motion of the "center of mass" of a couple
of trajectories $\omega, -\omega$. It is given by
\be
 \frac{z_{F, \omega}(t) + z_{F, -\omega}(t)}{2}= z_{F, \omega =0}
 =  {F(p)- p \over \kappa p}  \, .
\label{couple}
\ee
It is independent of $\omega$, as was the behavior of $\chi_F$.
In the relativistic case, $F=p$ for all $p$,
one recovers  $z=0$, the locus of the static event horizon. From
Eq. (\ref{couple}) we learn that the couple
leaves the near horizon region, i.e. $|z_F| = 1/\kappa $,
for $p \simeq p_c$. Therefore Eq. (\ref{peq}) applies only
until $p \simeq p_c$. This is the main effect of the
non-linear behavior of $F$: 
when the critical momentum is 
reached,
the center of mass starts to drift and this brings the pair of characteristics
outside the near horizon region, where the gradient of $v$ progressively vanishes.

Imagine indeed that the gradient of the flow smoothly decreases as
one leaves the near horizon region. Hence the acoustic metric
will become flat and $p$ will tend to a constant. In this steady regime,
far away from the horizon, 
positive $p$ characterizes $in$ vacuum for outgoing modes, irrespectively of the
dispersion relation $F$. Hence this proves that the $in$ vacuum is
still asymptotically
characterized by positive $p$ for $p \simeq p_c \gg \kappa$.
It should be emphasized that we did not refer to the formation
of the BH to characterize the $in$ vacuum. This is no longer necessary
because the blueshifting effect stops after a finite time and
the field configurations at the origin of HR leave the horizon region
and reach a flat region where a usual characterization of vacuum
by plane waves 
applies. Instead, in absence of dispersion,
the blueshifting effect only stops when the horizon forms.


The characterization of outgoing asymptotic quanta
is clearly unaffected by $F$ because it is done in a regime
where $p \simeq \kappa \ll p_c$.
The recovery of the unperturbed relativistic
behavior is illustrated by the fact that when $p \ll p_c$,
$z_F \propto p^2/p_c^2 \to 0$, i.e. $z_F$ tends
to the position of the horizon, as expected.

   From this analysis, we have learned that the
characterization of both $in$ modes and asymptotic $out$ modes
is unaffected by the introduction of $F$.
Hence the only source of modification of HR comes
from $B_F \neq 0$ in Eq. (\ref{wronchi2}).

\subsection{
The lessons}

Since several lessons have been drawn by
the analysis of HR in the presence of a non-linear DR
it is of value to present them synthetically.

{\bf I. HR is robust.}\\
The properties of HR are robust in the sense
that stationarity is fully preserved, whereas the modifications
of the spectrum are smaller or equal to $\kappa/ p_c\ll 1$.

{\bf II. The origin of the robustness of HR.}\\
It is also of value to 
specify why HR is so robust.
In the near horizon region, the modified modes
can be expressed as the relativistic modes times a
running wave function $\chi_F$ which is
independent of the frequency $\omega$, see Eq. (\ref{factoriz}).
Hence when there is no
backscattering and the three properties ({\bf A}, {\bf B } and {\bf C}
presented at the end of Section \ref{HRinacBH})
of the relativistic modes are unmodified, 
 no modifications of the Bogoliubov coefficients appear.
What governs the backscattered amplitude is a non-adiabatic
effect, like the Landau-Zener effect, which can even
be exponentially small, see Eq.(\ref{wronchi4}).

{\bf III. The exceptional feature of the relativistic propagation.}\\From
Eq. (\ref{Delta(z)}) we learn that the endless stretching found
in the relativistic case is an
exceptional feature resulting from the absence of any non-linearity in the DR.
This maintains the pair of characteristics
centered for all times on the event horizon $z=0$.
Instead, for all non-linear DR, 
 the departure from $\omega=cp$ as $p$ increases 
 induces a drift which sends the pair outside the near horizon region
where the gradient of velocity vanishes, hence where the blueshifting
effect ceases.

{\bf IV. Hints for Quantum Gravity.}\\
This last remark is an invitation to inquire about the
physical validity of the relativistic equation $\Omega^2 = c^2p^2$ 
to characterize field propagation near a horizon.
This equation results from the hypothesis that the bare actions
are locally Lorentz invariant, but also from a linearized treatment.
Without questioning the validity of Lorentz invariance
at a fundamental level, it is 
possible that radiative corrections evaluated
in the non-trivial near horizon metric will modify
1-loop Green functions
in such a way that the effective law governing their behavior
will be characterized by a new UV scale which shall act
like $p_c$ did it in the non-linear DR
we used. A first attempt to show that this
might be the case has been presented in \cite{parentaPeyresq}.
We refer to \cite{arteaga1,arteaga2}
for a more fundamental but less ambitious approach
to derive, from radiative corrections,
effective dispersion relations breaking the
(local) Lorentz invariance possessed by linearized field propagation.

\section{The phonons backreaction }

In the mean, the backreaction of the linearized phonons on the underlying fluid can be discussed
using the sound equations that govern the classical backreaction \cite{corto, lungo}, i.e. Eqs. (\ref{backeqstmunu1}, \ref{backeqstmunu2}) (see also \cite{bavili}) \bea & & \dot\rho_B + \vec\nabla (\rho_B\vec v_B)
+ \vec \nabla \sqrt{-g^B}\langle T^0_i\rangle =0\ , \label{backeqstmunu11}\\
& & \dot \psi_B + \frac{1}{2}\vec v_1^2 + \mu(\rho_B)
-\frac{1}{2}\frac{\rho_B}{c} \langle T\rangle  =0\ , \label{backeqstmunu22}\eea where
now $\langle T_{\mu\nu}\rangle$ are the quantum expectation values
of the pseudo energy momentum tensor operator of the phonon field.
Because of the analogy, the $\langle T_{\mu\nu}\rangle$ coincide
with the expectation values of the stress tensor for a minimally
coupled scalar field propagating in an effective acoustic
spacetime whose metric is $g^B_{\mu\nu}(\rho_B,\vec v_B)$ of Eq.
(\ref{gmunub}). \\ \noindent In the case of our interest, these
expectation values are to be taken in the $in$ vacuum, in which in
the remote past prior to the time dependent formation of the sonic
hole, the phonon field is in its vacuum state.\\ \noindent The
quantum version of the Euler equation, namely \be \dot{\vec v}_B +
(\vec v_B \cdot\vec\nabla)\vec v_B+\vec\nabla \mu(\rho_B)
-\frac{1}{2}\vec\nabla \left( \frac{\rho_B}{c}\langle T\rangle
\right)  =0\ee is the fluid analogue of the semiclassical Einstein
equations (see Appendix \ref{GrBack}) in gravity.

  The analogue of the
continuity equation (\ref{backeqstmunu1}) is the covariant
conservation of the stress tensor in Einstein's General
Relativity. \\ \noindent It is instructive to compare the role
played by the stress tensor in determining the dynamical evolution
of the system. Unlike semiclassical Einstein gravity, where all
components of $\langle T_{\mu\nu}\rangle $ enter the dynamical
equations (see Appendix \ref{GrBack}, Eq. (\ref{backeqs})), in hydrodynamics the dynamical backreaction is caused
just by the gradient of the trace density. \\ \noindent As they
stand, the expectation values of the minimally coupled scalar
field $\psi_1$ entering the backreaction equations
(\ref{backeqstmunu1}, \ref{backeqstmunu2}) are formally
ultraviolet divergent.~\footnote{Infrared divergences can be
simply eliminated by introducing a mass cutoff \cite{anderson}.}
\\ Here we are touching a very delicate point. In performing the
integrals over the frequencies involved in the expectation values
of $\langle T_{\mu\nu}\rangle $, one inevitably enters a regime
where the wavelength of the modes becomes comparable to the
intermolecular  
 separation. The approximation of the fluid as a
continuous  medium breaks down. High frequency modes see the
elementary constituents of the fluid and the dispersion relation
for these modes deviates significantly from the relativistic one
($\omega=c|\vec{p}|$) associated to the wave equation
(\ref{waveeqpsi1}) as we have discussed in section \ref{modmodeq}. We make the assumption that the contribution of
these modes to the expectation values is negligible. We shall give
later on 
 arguments in favor of this. Our working hypothesis
is therefore that the wave equation (\ref{waveeqpsi1}) holds for
all frequencies. Furthermore, we assume that the divergent expectation
values 
 are made 
 finite by adopting the standard covariant
renormalization scheme in curved space (see for example \cite{bd}), which, as such, preserves at
the quantum level the covariant conservation law of the classical
pseudo energy momentum tensor (i.e. $\langle T_\mu^{\
\nu}\rangle_{;\nu}=0 $). Within this framework one can eliminate
the divergent part of $\langle T_{\mu\nu}\rangle $.
\\ \noindent Unfortunately the finite part 
of $\langle
T_{\mu\nu}\rangle $ for an arbitrary metric
is not known and therefore the actual
acoustic metric, solution of the backreaction equations, can not
be determined.

\subsection{The dimensional reduction model}

Lower dimensional models are very useful in physics as they allow
explicit analytic solutions of the dynamical equations to be
found, providing at least a qualitative insight of the behavior of
real four dimensional systems. This attitude has been largely used
in studying quantum effects in black holes \cite{cghs, fabnav}.
The basic assumption  consists in replacing the four dimensional
quantum stress tensor by a two dimensional one associated to a two
dimensional minimally coupled scalar field propagating in the two
dimensional section of the physical four dimensional metric.  As
we know, this approximation neglects the backscattering of the
quantum field caused by the potential barrier.\\ We  follow the
same attitude for a qualitative insight in the quantum
backreaction occurring in a sonic black hole \cite{lungo,
corto}.\\ We shall focus on a supersonic flow through a de Laval
nozzle, which can be considered as one dimensional as discussed in
subsection \ref{lavalnozzle}. Under these assumptions, we can
write the classical equations of motion for the fluid as: \bea
\label{uppi}
&&A\dot{\rho}+\partial_z(A\rho v)=0\\
&&A\left( \dot{\psi}+\frac{1}{2}v^2+\mu(\rho)\right)=0
\label{uppo} \eea where $v=\partial_z \psi$ and we have also
assumed the time independence of the area $A$ of the transverse
section of the nozzle. Proceeding as in  section
\ref{classicalback}, we separate the perturbations from the mean
flow. Linear perturbations, i.e. sound waves or phonons at the
quantum level, satisfy the wave equation \be\label{wave2D}
\partial_a \left( \frac{A\rho}{c}\sqrt{-g_B^{(2)}}g^{B\ ab}\partial_b \psi_1\right)=0
\ee where \bea \label{twodmet}
  g^{B\ (2)}_{ab}=-\frac{\rho_B}{c}\left( \begin{array} {cc}
c^2-v_B^2&v_B\\
v_B &-1 \end{array} \right) \eea is the two dimensional ($t,z$)
section of the metric  $g_{\mu\nu}^B$ (\ref{gmunub})
and $a,b=t,z$. Note the invariance of Eq. (\ref{wave2D}) under Weyl transformations.\\
The two dimensional backreaction equations for the mean flow
$\rho_B, v_B$ read: 
 \bea \label{qev1}&&
A\dot{\rho_B}+\partial_z(A\rho_B
v_B)+\partial_z\left[-\frac{1}{c}\left(\langle
T_{tz}^{(2)}\rangle +v_B\langle T_{zz}^{(2)}\rangle\right)\right]=0\ , \\
\label{qev2}&& A\left(
\dot{\psi_B}+\frac{1}{2}v_B^2+\mu(\rho_B)\right)-\frac{1}{2}\langle
T^{(2)}\rangle=0. \eea Here $ T_{ab}^{(2)} =\frac{A\rho}{c}(\partial_a
\psi_1\partial_b \psi_1-\frac{1}{2}g^{(2)cd}\partial_c
\psi_1\partial_d \psi_1 g_{ab}^{(2)})$.\\
As seen from Eq. (\ref{wave2D}) the quantum field $\psi_1$ behaves
as a massless scalar field which is coupled not only to the two
dimensional metric $g^{B(2)}_{ab}$ but also to the "dilaton"
$\frac{A\rho}{c}$. This causes backscattering of the modes. The Polyakov
approximation consists in neglecting the backscattering. Eq. (\ref{wave2D}) is then
approximated as \be
\hat\Box^{(2)}\psi_1=0\
, \ee i.e. $\psi_1$ is treated as a two dimensional minimally
coupled massless scalar field. The reason of this simplification
is that even for the "exact" two dimensional theory of Eq.
(\ref{wave2D}) the exact $\langle T_{ab}\rangle $ is not known.
Only its trace is known being completely anomalous \cite{muk}. Given this assumption, the quantum
expectation values entering Eqs. (\ref{qev1}, \ref{qev2}) can
therefore be evaluated
using the methods developed in Appendix \ref{QuantumT}. \\
\noindent We shall use Eqs. (\ref{qev1}, \ref{qev2}) to find the
first order in $\hbar$ corrections to the classical fluid flow in
a de Laval nozzle which corresponds to a sonic black hole as
described in subsection \ref{abh}. To this end one writes the
dynamical variables $\rho_B$ and $v_B$ as \bea && \rho_B =
\rho_{cl} + \rho_q\ , \\ && v_B=v_{cl}+v_q \eea where $\rho_{cl}$,
$v_{cl}$ satisfy the classical equations (\ref{uppi}, \ref{uppo})
and $\rho_q$, $v_q$ are the corrections of order $\hbar$ induced
by the quantum fluctuations of the phonon field encoded in
$\langle T_{ab}^{(2)}\rangle $. \\ \noindent We can therefore
write the backreaction equations (\ref{qev1}, \ref{qev2}) as \bea
&&A\dot\rho_q + \partial_z\left[ A(\rho_q
v_{cl}+\rho_{cl}v_q)\right] -\partial_z \left[
-\frac{1}{c}(\langle T_{tz}^{(2)}\rangle + v_{cl}\langle
T_{zz}^{(2)}\rangle )\right]=0\ , \\ && A(\dot\psi_q + v_{cl}v_q
+\frac{c^2}{\rho_{cl}}\rho_q) - \frac{\langle T^{(2)}\rangle}{2}
=0\,\,\,\, . \eea Up to the order $O(\hbar)$, the $\langle
T_{ab}^{(2)}\rangle$ are computed on the classical solution
$\rho_{cl}$, $v_{cl}$, which
 are given in Eqs. (\ref{acbhse},
\ref{rhocl}), namely \bea && A=\frac{c}{|v_{cl}|}A_H
e^{(v_{cl}^2-c^2)/2c^2}\ ,\\ &&  \rho_{cl}=\rho_0
e^{-v_{cl}^2/2c^2} \ . \eea This configuration, as discussed in
subsection \ref{lavalnozzle}, should be regarded as the final
configuration of the fluid resulting from the time dependent
formation of the sonic hole.\\ \noindent
 We will be mainly interested in the region near
the sonic horizon $z=0$ where the profile of the nozzle can be
approximated as \be A=A_H + \beta z^2\ . \ee For the sake of
simplicity, the suffix "cl" will be omitted in the following.
\\ \noindent Standard covariant renormalization in curved space allows, under
the previous assumption, the expectation values of the phonon
fields in Eqs. (\ref{qev1}, \ref{qev2}) to be evaluated (see
Appendix \ref{QuantumT}). For the trace term we have simply (trace
anomaly) \cite{duff} \be \langle T^{(2)}\rangle =
\frac{\hbar}{24\pi}R^{(2)}\ , \ee whereas in the continuity
equation (\ref{qev1}) we can write \be \frac{1}{c}(\langle
T_{tz}^{(2)}\rangle + v\langle T_{zz}^{(2)}\rangle ) =
\frac{\langle T_{++}\rangle }{(c-v)^2} - \frac{\langle
T_{--}\rangle}{(c+v)^2}\ , \ee where we have used the  null
(retarded and advanced) coordinates (\ref{xpmnull}) \be x^\pm =
c\left( t\pm
\int \frac{dz}{c\mp v}\right)\ . \ee 
In our double null coordinates the components of the stress tensor
are (see for example \cite{bd})\be \langle T_{\pm\pm}\rangle
=-\frac{\hbar}{12\pi}C^{1/2}C^{-1/2}_{,\pm\pm}+ t_{\pm}\ , \ee The
conformal factor $C$ of the two dimensional metric (\ref{twodmet})
reads \be C=\frac{\rho}{c^3}(c^2-v^2)\ee and $t_{\pm}$ are
functions of $x^{\pm}$ respectively, which depend on the choice of
the quantum state in which the expectation values are to be taken.
\\ \noindent Near the horizon ($x^-\to
+\infty$) the $t_{\pm}$ can be approximated with the ones 
characterizing the $in$ state at late time \cite{unruh76}, namely  
no incoming radiation in the past:
\be t_+=0\ ,
\ee 
and thermal radiation of phonons at the Unruh temperature in the future:
\be
t_-=\frac{\hbar k^2}{48\pi c^4}\ . \ee This latter
represents the Hawking radiation flowing at late time at $z\to
\infty$. \\ \noindent For our configuration the surface gravity
$k$ reads \be k=c^2\k=c^2\sqrt{\frac{\beta}{A_H}}\ . \ee Expanding
the classical background near the horizon $z=0$ to evaluate the
$\langle T_{ab}^{(2)}\rangle$, one eventually arrives at the
following solution of Eqs. (\ref{qev1}, \ref{qev2}) for the
quantum corrected velocity \be \label{vb}v_B=v+v_q=-c+c\k z
-\frac{c}{6}\k^2 z^2+\epsilon (b_1 + c_1\k z)\k t\ , \ee where
$\epsilon \equiv \frac{\hbar}{A_H^2 \rho_0 e^{-1/2}c}$ is the
dimensionless expansion parameter and $b_1=9\gamma/2$,
$c_1=-304\gamma/15$, $\gamma\equiv \k c^2A_H/24\pi $, and  \be
\label{rhobi}\rho_B=\rho+\epsilon\rho_q= \rho_0 e^{-1/2}\left(
1+\k z -\k^2z^2/6 +\epsilon \frac{1}{c^2}(\alpha + \delta \k
z)t\right)\ , \ee where $\alpha=-a_2+c\k b_1$, $\delta=-b_2-a_2
+c\k c_1$, with $a_2=1189\gamma c\k/120$, $b_2=-161817\gamma
c\kappa/120$. The solution is valid for $\k z\ll 1$ (near horizon
approximation) and $c\k t\ll 1$. So equations (\ref{vb},
\ref{rhobi}) give just a hint on how the backreaction starts near the
sonic horizon. \\ \noindent The boundary condition chosen to
derive Eqs. (\ref{vb}, \ref{rhobi})  is that, at some given time
(say $t=0$), the evaporation is switched on starting from the
classical configuration ($\rho, v$): i.e. $v_q(t=0,z)=0$,
$\rho_q(t=0,z)=0$.
\\ \noindent Inspection of Eq. (\ref{vb}) reveals that the net
effect of backreaction is to slow down the fluid ($v_q>0$ for $\k
z\ll 1$). This goes with a decrease of the density ($\rho_q<0$) in
the same limit. The same equation allows to determine the
evolution of the acoustic horizon, defined as the locus where $v_B=-c$. We find
\be z_h =- \frac{\epsilon b_1 t}{c}\ , \ee i.e. the horizon is
moving to the left: the hypersonic region shrinks in size. The
coefficient $b_1$, which determines the quantum correction to the
velocity field, and hence the evolution of the horizon, is
proportional to the gradient of the trace calculated at $t=0$,
$z=0$, as it can be seen by taking the gradient   
 of Eq. (\ref{qev2}) in
this limit. This should be compared to the gravitational black
hole case where the evolution of the horizon is determined by the
energy flux ($\dot M \propto \langle T^r_{\ t}\rangle $ in
spherical symmetry). While in the latter case Hawking radiation
occurs at the expense of the gravitational energy of the black
hole (the mass $M$), 
the  phonons emission takes away
kinetic energy from the fluid.
\\ \noindent
Considering the evolution of the acoustic black hole as a sequence
of quasi stationary configurations one can find the correction to
the emission temperature \be T=\left. \frac{\hbar}{2\pi}\frac{\partial
v}{\partial z}\right|_{z_H} = \frac{\hbar c}{2\pi}\k \left[ 1-
\frac{563}{720\pi}\epsilon \k^3 cA_H t\right]\ . \ee As time
evolves, the temperature of the emitted phonons decreases. \\
\noindent As discussed in Appendix \ref{GrBack},  the
opposite  happens for a gravitational black hole of mass $M$. Being $T \sim 1/M$, as the mass decreases because of
Hawking evaporation, the temperature grows causing the black hole
to evaporate completely in a time of order $M_0^{3}$ ($M_0$ is
the initial mass of the hole). Instead, if the black hole possesses
a conserved $U(1)$ charge $Q$, for $M \stackrel{>}{\sim} |Q|$ a
temperature decrease is associated to the mass evaporation, which
stops when it is reduced to $M=|Q|$  (the so called extremal black
hole) for which the Hawking temperature is zero. This ground state
is approached in an infinite time. Acoustic black holes seem
therefore to behave like near extremal charged black holes.

\subsection{Discussion}

An important question which remains open is to what extent the 
above
results 
depend on the two assumptions we made, namely,
first, the use of a relativistic
 dispersion relation 
 which
ignores 
the
short distance corrections due to the molecular structure of the fluid,
and second, the covariant regularization scheme used to subtract ultraviolet divergences.
These two aspects are deeply connected.
For the Polyakov theory that we have used in the 2D backreaction equations (\ref{qev1},
\ref{qev2}), Jacobson \cite{ted93} has argued that, within a covariant regularization scheme,
no significant deviation from the usual expression for the trace anomaly and the components
of $\langle T_{ab}^{(2)}\rangle$ are expected if one introduces a cutoff at high frequencies.
For the hydrodynamical system we have considered, covariance is a symmetry of the
phonons low energy effective theory only, which is broken
at short distance.
Hence non covariant terms depending on the microscopic physics are expected to show up in the effective action,
and these are crucial for a correct description of the unperturbed quantum vacuum of the fluid.
However the expectation values $\langle T_{\mu\nu}\rangle$ entering the backreaction equations
(\ref{backeqstmunu11}, \ref{backeqstmunu22})
do not represent the energy momentum of the fluid quantum vacuum.
They describe instead the perturbation of the stationary vacuum (whose energy is strictly
zero \cite{volovik}) induced by inhomogeneities and by the time dependent formation
of the sonic hole which triggers the phonons emission. Here we have assumed that these deviations can be computed
within the low energy theory. This situation is not unusual.
Casimir effects are well known examples of vacuum disturbances caused by the presence of boundaries.
It happens that the Casimir energy is often (but not always,
see G. Volovik in Ref. \cite{libro})
independent on the microscopic physics and can be calculated within the framework of the low
energy theory. As shown by Bachmann and Kempf \cite{kempf}, deviations caused by the modified dispersion relation are suppressed by positive powers of $d_c/L$, where $d_c$ is the interatomic distance of the material constituting the walls and $L$ the separation between the walls.
This happens because, while low frequency modes are reflected by the
boundaries, for the high energy
ones the wall is transparent.  They produce a divergent contribution to the vacuum energy which
is canceled by a proper regularization scheme and does not affect the finite result.
Similarly, in \cite{pomeau} it is observed that, although in a Bose-Einstein condensate quantum fluctuations have a non-linear dispersion relation, the Casimir energy is given by the long wavelength modes, which therefore behave as massless scalar fields.
We have assumed
that a similar decoupling happens for the acoustic black hole.
The check of our hypothesis would require an analysis of the quantum system within
the microscopic theory which takes into account
the time dependent non homogeneous formation of the sonic hole.
This is for the moment beyond computational capability.
Anyway, as it has been shown in section \ref{modmodeq}, modifications of the dispersion relations, to take into account
short-distance behavior of the high-energy
modes, basically do not affect the spectrum of the emitted phonons.
This is not a proof that observables like $\langle
T_{\mu\nu}\rangle$ are also unaffected by short-distance physics. However, 
keeping in mind what
$\langle T_{\mu\nu}\rangle$ does really represent (i.e. deviations in energy and momentum caused by the inhomogeneity of the flow), and the indications coming from the Casimir effect, one expects the contribution coming from the relativistic low enegy theory to be the most significant in driving the backreaction.

\section{Conclusions and open questions}

In section 2 we show how a relativistic wave equation
emerges when studying the linear perturbations
of an Eulerian fluid which possesses a Galilean invariance.
As mentioned in the footnote 3, one gets a similar
wave equation when considering relativistic fluids.
In both cases, the wave equation defines an "acoustic metric",
i.e. a Lorentzian four-dimensional metric
which is curved when the fluid's flow is non-homogeneous
and/or non-stationary. Similar wave equations appear in many non-linear systems,
such as non-linear electrodynamics \cite{novello},
and di-electric media \cite{leonhardt}.
It would be interesting to identify what are the
physical ingredients (such as local equilibrium)
necessary to obtain a Lorentzian wave equation
for linear perturbations.

We conclude section 2 by analyzing the first backreaction
(non-linear) effects, namely how the mean fluid's flow
is affected by the energy carried by an ensemble of sound waves.
We show that these effects are governed by the pseudo energy-momentum
tensor associated with the four-dimensional metric, see Eqs.
(\ref{backeqstmunu1}, \ref{backeqstmunu2}). It would be
also interesting to see if this result extends to other physical
situations wherein linear perturbations obey a Lorentzian wave equation.

In sections 3 and \ref{HRinacBH},
we review the essential features of quantum field
theory in curved spacetime, focussing on the case of interest:
the stationary curved metric describing a (static) black hole
geometry. We emphasize that the stationary modes associated
with asymptotic quanta are singular on the horizon,
while the quantum vacuum state is regular there. This potential conflict
is resolved by the existence of Hawking Radiation, which is
encoded in the non-trivial Bogoliubov coefficients relating the
singular  asymptotic modes to the regular modes
describing the vacuum state, see Eq. (\ref{phi.in(z)}).
We have phrased the derivation in the $p$-representation
to prepare the analysis of HR in the presence of dispersion,
which appears at short distances in materials.

In section \ref{modmodeq} we explain
why the properties of HR, namely stationarity and thermality,
are not affected when introducing a non-linear dispersion relation,
in spite of the facts that Lorentz invariance is broken at short distances,
and that the field propagation is dramatically modified
in the near horizon region. From this analysis we learn that
the properties of HR are robust, but also that the near horizon
behaviour of the modes in the relativistic case is exceptional
(unstable) in the following sense.
Any non-linearity in the dispersion relation will
indeed erase their singular behavior because the
blue-shifting effect obtained in a backwards in time extrapolation
stops when the UV scale is reached, see Eq. (\ref{couple}).
Possible implications of this result for Quantum Gravity are
briefly discussed.

In section 6 we compute the secular effect on the fluid's flow induced bybHawking radiation. The
expectation values 
$\langle T_{\mu\nu}\rangle$ of the pseudo energy-momentum tensor of the
phonons drive this backreaction. Their computation depends in
principle on the UV behavior of the dispersion relation. However,
there are strong indications that the most significative
contribution to $\langle T_{\mu\nu}\rangle$ comes from the low
energy theory (i.e. relativistic dispersion relation). Given this
the evaporation of a sonic black hole seems characterized by a
slowing down of the fluid associated to a decrease of the
temperature of the emitted phonons. \\ In this analysis the
starting point was the action (\ref{hydroaction}) and the fields
$\rho$ and $\psi$ were considered as fundamental. In many
condensed matter systems, however, the hydrodynamical action
represents just an approximation of the real action, e.g. in  Bose-Einstein condensates. For these systems, the  backreaction problem should be addressed at the level of the real action. The correction to this more rigorous approach with respect to the
hydrodynamical one developed so far is matter for future
investigations \cite{schutz}.\\

\noindent {\bf Acknowledgments}:
The 
authors thank
S. Liberati, M. Rinaldi and M. Visser
for
useful comments and suggestions. S. F. also thanks Centro Enrico Fermi
for supporting her research.
\section*{Appendix}
\begin{appendix}
\section{General Relativity}\label{GR}

General Relativity is a geometric theory of gravity. According to
it the gravitational interaction manifests itself
as curvature of
the spacetime manifold. The gravitational field is described by
ten potentials which are the independent components of the
geometric tensor $g_{\mu\nu}$, giving the "distance" in
spacetime between neighbouring events \be ds^2=g_{\mu\nu}dx^\mu
dx^\nu \ . \ee According to the principle of equivalence, the
effects of gravity can be locally eliminated  going to the local
inertial coordinate system in which the metric tensor assumes the
Minkowski form $\eta_{\mu\nu}=diag(-1,1,1,1)$ and the connection
coefficients \be \Gamma^\alpha_{\mu\nu}=\frac{1}{2}g^{\alpha\beta}\left(g_{\mu\beta ,\nu}+ g_{\nu\beta
,\mu} -g_{\mu\nu ,\beta}\right) \ee vanish. \\ \noindent Gravity,
however, shows up  as relative acceleration between
neighbouring points causing tidal distortions measured by the
Riemann tensor of the spacetime \be R^\mu_{\nu\alpha\beta}=\Gamma^\mu_{\nu\beta ,\alpha}-\Gamma^\mu_{\nu\alpha ,\beta}+
\Gamma^\mu_{\gamma\alpha}\Gamma^\gamma_{\nu\beta}-
\Gamma^\mu_{\gamma\beta}\Gamma^\gamma_{\nu\alpha}\ .\ee In
absence of gravitation spacetime is flat and the Riemann tensor
identically vanishes (and vice versa). \\ \noindent The sources of
the gravitational field are the distributions of mass-energy of
the matter and gauge fields encoded in their energy momentum tensor
$T_{\mu\nu}$. This determines the gravitational potentials, up to
coordinate transformations, via the Einstein equations (here and in the following we take $G=c=1$)
\be\label{einsteqs} R_{\mu\nu}-\frac{1}{2}Rg_{\mu\nu}=8\pi
T_{\mu\nu}\ . \ee Here $R_{\mu\nu}$ is the Ricci tensor \be
R_{\mu\nu}=R^\alpha_{\mu\alpha\nu}\ee and $R$ the Ricci scalar \be
R=g^{\mu\nu}R_{\mu\nu}\ .\ee
 Because of the Bianchi identities the stress tensor
is covariantly conserved \be {T_\mu^{\ \nu}}_{;\nu}=0\ , \ee
where $;$ means covariant derivative with respect to the metric $g_{\mu\nu}$. This
describes how energy and momentum  are exchanged between  matter-gauge
fields and the gravitational field. \\ \noindent In the vacuum,
$T_{\mu\nu}=0$ and Einstein equations simplify to
\be\label{einsteqsvac} R_{\mu\nu}=0\ .\ee Once the gravitational
field is determined, by Eqs. (\ref{einsteqs}) or outside the
matter energy distribution by Eqs. (\ref{einsteqsvac}), one can
analyze the motion of test particles. These move along the
geodesics of the curved spacetime, determined by \be\label{geodesics} \frac{d^2
x^\mu}{d\tau^2}+
\Gamma^\mu_{\alpha\beta}\frac{dx^\alpha}{d\tau}\frac{dx^\beta}{d\tau}=0\ee
where $\tau$ is an affine parameter along the geodesic, which, for massive particles, can be
chosen as proper time. In term of the tangent vector $K^\alpha=\frac{dx^\alpha}{d\tau}$ the geodesic equation can also be given as \be\label{KK}
K_{\alpha;\beta}K^\beta=0\ .\ee  Geodesics of interest
are either timelike (with tangent vector
normalized to $K_\mu K^\mu =-1$) in
the case of massive particles, or null ($K_\mu K^\mu =0$) in the
case of massless particles (for example photons).

\section{Gravitational waves}
\label{gw}

Gravitational waves are curvature ripples rolling across the
spacetime. We shall briefly outline how to derive the propagation
equations and the characteristic features of gravitational waves.
A more complete and detailed discussion can be found in
\cite{mtw}. \\ \noindent We shall consider gravitational waves
propagating through a vacuum spacetime. Let $R$ be the typical
radius of curvature of the background, $\lambda/2\pi$ and $A$
 the typical reduced wavelength and amplitude of the
waves respectively. We assume $A\ll 1$ and $\lambda/2\pi\ll R$. The spacetime
metric $g_{\mu\nu}$ is expanded in a  "background" $g_{\mu\nu}^{(B)}$, a smooth part which included the backreaction of the waves,
plus perturbation $h_{\mu\nu}$, i.e. \be
g_{\mu\nu}=g_{\mu\nu}^{(B)}+h_{\mu\nu} \ .\ee The Ricci tensor can
be expanded as \be
R_{\mu\nu}=R_{\mu\nu}^{(B)}+R_{\mu\nu}^{(1)}(h)+R_{\mu\nu}^{(2)}(h)+O\left(\frac{A^2}{\lambda^2}\right)
\ee where $R_{\mu\nu}^{(B)}$ is the Ricci tensor constructed from
$g_{\mu\nu}^{(B)}$, while
\bea R_{\mu\nu}^{(1)}&\equiv&\frac{1}{2}\left( -h_{;\mu\nu}
-h_{\mu\nu;\alpha }^{\ \ \ \ \alpha}+h_{\alpha\mu;\nu }^{\ \ \ \ \alpha}
+h_{\alpha\nu;\mu }^{\ \ \ \ \alpha}\right)\\
R_{\mu\nu}^{(2)}&\equiv&\frac{1}{2}\left[ \frac{1}{2}h_{\alpha\beta;\mu}
h^{\alpha \beta}_{\ \ \ ;\nu}+h^{\alpha \beta}\left(  h_{\alpha\beta;\mu\nu}+h_{\mu\nu;\alpha\beta}
-h_{\alpha\mu;\nu\beta} -h_{\alpha\nu;\mu\beta} \right)+\right. \nonumber \\
&& \left. h_\nu^{\alpha;\beta}\left( h_{\alpha\mu;\beta}
-h_{\beta\mu;\alpha} \right)-\left( h^{\alpha \beta}_{\ \ \
;\beta}-\frac{1}{2}h^{;\alpha}\right)\left(
h_{\alpha\mu;\nu}+h_{\alpha\nu;\mu}-h_{\mu\nu;\alpha} \right)
\right].\;\;\; \ \ \ \ \ \eea Here and in the following indices are
raised and lowered with $g_{\mu\nu}^{(B)}$, $h\equiv
g^{\alpha\beta (B)}h_{\alpha\beta}$ and $";"$ indicates
covariant derivative with respect to $g_{\mu\nu}^{(B)}$.\\
\noindent The vacuum field equations require $R_{\mu\nu}=0$. The
linear part in the amplitude $A$, i.e. $R_{\mu\nu}^{(1)}$, must
vanish since the distorting effect of the wave on  the background is
a non linear phenomenon, so $R_{\mu\nu}^{(B)}$ cannot be linear in
$A$. Hence \be \label{propeqgw} R_{\mu\nu}^{(1)}(h_{\mu\nu})=0 \ .
\ee  These are the propagation equations for the gravitational waves.\\
\noindent The rest  of $R_{\mu\nu}$ can be written as \be
R_{\mu\nu}^{(B)}+\langle R_{\mu\nu}^{(2)}(h)\rangle
+O\left(\frac{A^2}{\lambda^2}\right) =0\ee averaging over several
wavelengths. We can rewrite this in a suggestive form  \be
R_{\mu\nu}^{(B)}-\frac{1}{2} R^{(B)}g_{\mu\nu}^ {(B)}=8\pi
T_{\mu\nu}^{(gw)}\ee where \be\label{Tmunu.gw}
T_{\mu\nu}^{(gw)}=-\frac{1}{8\pi}\left( \langle
R_{\mu\nu}^{(2)}(h)\rangle -\frac{1}{2}g_{\mu\nu}^ {(B)} \langle
R^{(2)}(h)\rangle \right)\ee is the effective stress
tensor for the gravitational waves. \\ \noindent The propagation
equations (\ref{propeqgw}) can be further simplified in terms of
\be \bar{h}_{\mu\nu}=h_{\mu\nu}-\frac{1}{2}h g_{\mu\nu}^ {(B)}\
.\ee Reparametrization invariance can be used to enforce the
"Lorenz gauge condition" \be ({\bar{h}_\mu^\alpha})_{;\alpha}=0 \
.\ee In this gauge, the vacuum propagation equations for the
gravitational waves becomes \be \Box \bar{h}_{\mu\nu}+2
R_{\alpha\mu\beta\nu}^{(B)}\bar{h}^{\alpha\beta}=0 \ .\ee
In the geometric optics approximation, one assumes  $\bar{h}_{\mu\nu}$
to have an amplitude $A_{\mu\nu}$ that varies slowly (on scale $l\ll R$) and
a phase which varies rapidly, namely \be \bar{h}_{\mu\nu}=\Re
(A_{\mu\nu}+..)e^{i\theta}  \ .\ee One can show that the rays (the
curves perpendicular to the surface of constant phases) are null
geodesics, i.e. \bea K_\alpha K^\alpha&=&0\ , \\ K_{\alpha;\beta}
K^\beta&=&0\ ,  \eea where $K_\alpha\equiv \theta_{,\alpha}$.
Being $K^\alpha$ a null vector, gravitational waves propagate at
the velocity of light.

\section{Black holes}
\label{black.holes}

\begin{figure}
\includegraphics[angle=270,width=5.0in,clip]{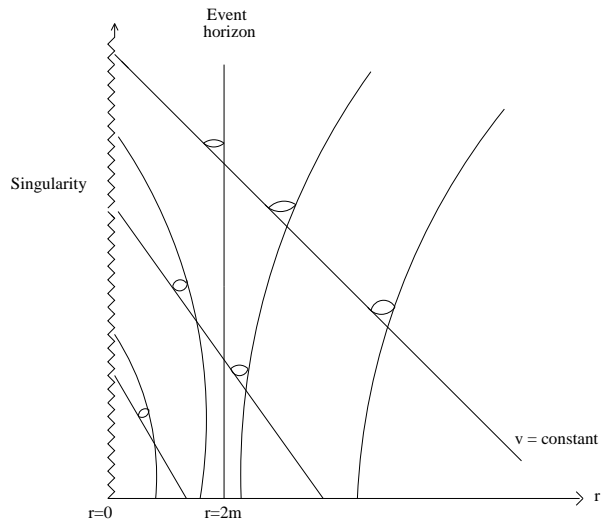}
\caption{Diagram representing the radial geodesics tangent to the light cones
in the spacetime of a Schwarzschild black hole. On the vertical axis we have
increasing times, while on the orizontal one we have the radial coordinate $r$
(two angular coordinates have been neglected). In $r=0$ there is the curvature  singularity, while  the horizon is located at  
$r=2M$. The region 
$r<2M$ is 
the gravitational black hole. The light cones are increasingly
distorted when approaching the horizon. Once the horizon is
crossed, they all point towards the singularity. This shows why
nothing can ever escape from the black hole (see also Fig.
\ref{fig6}).}
\label{fig5}
\end{figure}

\begin{figure}
\includegraphics[angle=270,width=5.0in,clip]{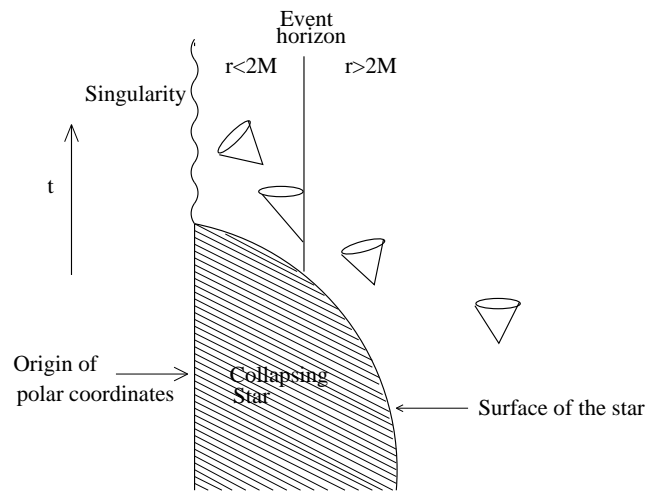}
\caption{Representation of light cones outside  a collapsing star
(shaded region) forming  a black hole. Before the horizon formation,
and far away from the star surface, the light cones are
only slightly distorted. After the horizon formation, 
the regions $r>2M$ and $r<2M$ (the black hole) are causally disconnected: no signal from the $r<2M$ region can ever reach  the outer region.}
\label{fig6}
\end{figure}

Since gravity couples to energy-momentum it affects not only the
motion of massive particles, but also of light. In the presence of
a gravitational field, light does not follow a rectilinear motion,
but it moves along the null geodesics (\ref{geodesics}) of the
curved spacetime. The deflection caused by the gravitational field
on the motion of light is very small in "weak" 
gravitational field like the one of our Solar System. However,
there are astrophysical systems where this deflection is so strong
that light cannot escape from some compact region of space that
becomes causally disconnected from the rest of the Universe. This
region is called black hole, the most spectacular consequence of
Einstein's General Relativity. In a black hole, the gravitational
field is so strong that light is dragged back instead of escaping
from it.\footnote{Since nothing can propagate faster than light,
the same thing happens to any matter field.} The simplest example
of a black hole is the one based on the Schwarzschild solution to
the Einstein equations. This is the unique solution of vacuum
Einstein Eqs. (\ref{einsteqsvac}) under the hypothesis of
spherical symmetry (according to the Birkhoff theorem
\cite{hawell}).
\\
\noindent In the so-called Schwarzschild coordinates the
 line element reads \be\label{schw}
ds^2=-(1-2M/r)dt^2+(1-2M/r)^{-1}dr^2 +r^2(d\theta^2 + \sin^2\theta
d\varphi^2)\ .\ee This metric describes the gravitational field of
a spherical body of mass $M$. It is asymptotically flat (i.e.
$g_{\mu\nu}\to \eta_{\mu\nu}$ for $r\to \infty$) and has the
correct Newtonian limit. 
The metric presents singularities at $r=2M$ and $r=0$. The latter
is a curvature singularity (i.e. the invariants constructed from the
Riemann tensor diverge), which signals a breakdown of the
theory there. On the other hand, the singularity at $r=2M$  is
just an artifact of the coordinates system used and can be
eliminated by a coordinate transformation. For example, using
the advanced or ingoing Eddington-Finkelstein coordinate
\cite{eddfin} \be v=t+r^*\ ,\ee where $r^*=\int
\frac{dr}{(1-2M/r)}=r+2M\ln|r/2M -1|$, the Schwarzschild line
element can be put in the following form \be\label{advedfink}
ds^2=-(1-2M/r)dv^2 + 2dvdr + r^2d\Omega^2\ ,\ee
where $d\Omega^2=d\theta^2+\sin^2\theta d\varphi^2$ is the line element of a unit two-dimensional sphere.\\ Due to the cross
term $drdv$ in these coordinates the surface $r=2M$ is 
regular. It is a null surface with special properties, as we shall
see in short. The radial null geodesics ($\theta=const.,\
\varphi=const.$) of the metric (\ref{advedfink}) are found by
setting $ds^2=0$; one finds \be \label{ingeoef}v=const. \ , \ee
which describe the motion of ingoing spherical light fronts, and
\be \label{outgeoef}\frac{dr}{dv}=\frac{1}{2}(1-2M/r)\ee for
outgoing light. A plot of both families of geodesics is given in
Fig. \ref{fig5}. Ingoing ones are depicted as straight lines.\\
\noindent
Inspection of Eq. (\ref{outgeoef}) clearly indicates that outgoing
rays manage to move outwards only for $r>2M$ (for which
$\frac{dr}{dv}>0$), whereas, for $r<2M$, they move
inwards ($\frac{dr}{dv}<0$)! For $r=2M$ they are stuck there
forever. \\ \noindent Therefore, light and any  matter field
cannot escape from the region $r<2M$, which represents the black
hole. The boundary of this region (i.e. $r=2M$) is the horizon, which can be
crossed only from outside to inside. The
spacetime given by the line element in (\ref{advedfink}), and
whose  structure is depicted in Fig. \ref{fig5}, describes,
according to relativistic astrophysics, the gravitational field
exterior to a sufficiently massive spherical star
($M\stackrel{>}{\sim} 3M_{sun}$) at the end of its thermonuclear
evolution, when the internal pressure is no longer able to prevent
the star collapsing because of  its own gravity \cite{htww}. Then the star undergoes
a catastrophic gravitational collapse that leads to the formation
of a black hole, as shown schematically in Fig. \ref{fig6}.
Here, also  the causal light cones, whose generators are the
null geodesics (\ref{ingeoef}, \ref{outgeoef}), are depicted. For $r<2M$ light,
and therefore any kind of matter, is forced to move inwards towards
the singularity. Hence, once the star surface crosses the  horizon,
it becomes disconnected with respect to the rest of the Universe.
After this, informations on the star cannot any longer be retrieved
by observers safely  outside the horizon. The only information left about the star structure is its gravitational mass $M$, which is also the only
parameter in the line element (\ref{advedfink}) \cite{isr}.
This is the so called no-hair theorem \cite{mtw}. \\
\noindent In practice signals from the surface of the star
disappear before the formation of the horizon. Let one consider the emission of radiation with
frequency $\nu_0$ with respect to  the free falling frame. If the emission occurs at some point with coordinate
$r$ ($r>2M$), then the frequency $\nu_{\infty}$ 
measured at $r\gg 2M$ is \be \nu_{\infty}=\nu_0(1-\frac{2M}{r})\
.\ee This is a manifestation of the  redshift
suffered by radiation climbing  the gravitational field of the
star. For $r\to 2M$, $\nu_{\infty}\to 0$  exponentially, namely
\be\label{redshiftgrav} \nu_{\infty}\sim \nu_0e^{-ku}\ ,\ee where
$u$ is a retarded time defined as \be u=v-2r^*=t-r^* \ee which is
constant along outgoing radial null geodesics (see Eq.
(\ref{outgeoef})): $u\to \infty$ as $r\to 2M$. $k$ is the so-called surface gravity at the black hole horizon. For a
Schwarzschild black hole \be \label{surgragrav}k=\frac{1}{4M}\
.\ee The surface gravity measures the strength of the exponential
falloff. It is formally defined through the formula \cite{wald}
\be  (\chi^\nu\chi_\nu )_{;\mu}=-2k\chi_\mu\ , \ee where
$\chi^\mu=(\frac{\partial}{\partial t})^\mu$ is the timelike
Killing vector field  whose norm vanishes at the horizon.
This quantity, as we 
have seen,
plays a central role
in quantum effects. 
\\ \noindent 
The so-called Painlev\'e-Gullstrand 
coordinates \cite{paingull} (see the more recent \cite{poisson}) provide another regular coordinate system across the future horizon. Moreover they naturally appear in the description of acoustic black holes.
In ingoing Painlev\'e-Gullstrand coordinates, the line element
reads
\be\label{pg} ds^2=-\left(1-\frac{2m}{r}\right) dt^2+
\sqrt{\frac{2M}{r}}dr dt +dr^2+r^2(d\theta^2+\sin^2\theta
d\phi^2)\ . \ee
The Painlev\'e-Gullstrand coordinates are related to the usual
Schwarzschild coordinates by the 
transformation \beq dt_{PG}=dt_S + \frac{\sqrt{2M/r}}{1-2M/r}dr\
. \eeq This is defined in   a coordinate patch that covers the
asymptotic region ($r\rightarrow \infty$) together with the region containing the future horizon and the black hole  singularity.\\ \noindent The
Painlev\'e-Gullstrand line element exhibits some special
features. For instance,  the constant time spatial slices are
 flat, with zero curvature. The spacetime
curvature of the Schwarzschild geometry is encoded into the
time-time and time-space components of the metric.

As we have seen, spherically symmetric vacuum (i.e. satisfying
$R_{\mu\nu}=0$) black holes are characterized by  the
gravitational mass $M$ only.  However, there  also  exist charged
black holes, characterized, in addition, by an Abelian charge $Q$
(which can be either electric or magnetic). These are solutions of
the full Einstein equations (\ref{einsteqs}) in which the source
(i.e. $T_{\mu\nu}$) is the stress energy tensor for the
electromagnetic field. Usually the latter is expressed  in terms
of the four- potential $A_{\mu}$ \be
F_{\mu\nu}=\partial_{\mu}A_{\nu}-\partial_{\nu}A_{\mu}\ ,\ee and
satisfies the curved version of the Maxwell equations \be
{F^{\mu\nu}}_{;\nu}=0\ ,\ee whereas the gravitational field is
obtained by solving \be R_{\mu\nu}-\frac{1}{2}Rg_{\mu\nu}=8\pi
(F_{\mu\sigma}F_\nu^\sigma -
\frac{g_{\mu\nu}}{4}F_{\rho\sigma}F^{\rho\sigma})\ .\ee These
combined Einstein-Maxwell equations admit a unique solution with
spherical symmetry, namely  the Reissner-Nordstr\"om
\cite{reinor}, with line element \be \label{reissnord}
ds^2=-(1-2M/r+Q^2/r^2)dt^2
+(1-2M/r+Q^2/r^2)^{-1}dr^2+r^2d\Omega^2\ ,\ee while the only
non-vanishing component of the Maxwell tensor is \be
\label{electr} F_{rt}=\frac{Q}{4\pi r^2}\ ,\ee for the electric
case (radial electric field), or \be
\label{magnet}F_{\theta\varphi}=\frac{Q}{4\pi r^2}\ ,\ee in the
magnetic case (radial magnetic field). Eqs. (\ref{reissnord},
\ref{magnet}) describe the gravitational and electromagnetic
fields respectively outside a spherical body of charge $Q$ and
mass $M$. For $M>|Q|$, the Reissner-Nordstr\"om metric Eq.
(\ref{reissnord}), besides the usual curvature singularity at
$r=0$,  has two coordinate singularities at \be \label{horrntwo}
r_{\pm}=M\pm\sqrt{M^2-Q^2}\ .\ee  The singularities at $r_{\pm}$
coincide when
$M=|Q|$, while no coordinates singularity is present for $M<|Q|$. \\
\noindent As before,  these coordinate singularities are not
physical, they are artifacts of the coordinates system and the
spacetime metric can be regularly extended across them. When
$M>|Q|$ the solution describes a black hole with  horizon
located at $r_+$. For $r_-<r<r_+$ light and matter are forced to
move inwards. However, as the inner horizon $r_-$ is crossed, geodesics can avoid the curvature
singularity ($r=0$) as they bounce  back and propagate towards
an asymptotically flat region, which is different from the one in which they have been generated \cite{hawell}. The bounce is caused by the electromagnetic
contribution, namely the $Q^2/r^2$ term present in the components of the metric, which causes the
gravitational field to be repulsive at small values of $r$.\\
\noindent The  horizon at $r_+$ is an infinite redshift
surface characterized by a surface gravity \be \label{horrn}
k_{rn}=\frac{\sqrt{M^2-Q^2}}{r_+^2}\ . \ee
 The inner horizon $r_-$ is an infinite
blueshift surface, which, unlike the event horizon at $r_+$, is
unstable against perturbations . The infinite blueshift causes
perturbations to grow and develop a curvature singularity which
prevents any continuation across $r_-$. This phenomenon is called  mass
inflation \cite{poiisr}. \\ \noindent For $M=|Q|$, the two roots of
Eq. (\ref{horrntwo}) coincide and one has a so-called extremal black hole
with a unique horizon at \be r_h=M\ , \ee characterized by vanishing
surface gravity. \\ \noindent Finally,  the Reissner-Nordstr\"om
solution for $M<|Q|$ has no horizons hiding the curvature
singularity at $r=0$. It describes a so-called "naked singularity". The
cosmic censorship hypothesis forbids the formation of such an
object starting from smooth initial conditions. However,  no explicit proof of this conjecture proposed by Penrose in
1969 \cite{penrose} is known yet.

\section{The backreaction in gravitational black holes}
\label{GrBack} Spacetime curvature has a non trivial effect on
quantum fields propagating on it, inducing vacuum polarization and
particles creation. 
In the mean, these are described by the quantum expectation
values $\langle T_{\mu\nu}\rangle$ of the stress energy tensor
operator for quantum matter and gauge fields. In principle,
$\langle T_{\mu\nu}\rangle$ also contain the one-loop gravitons
contribution, which is the expectation value of the quantum
version of the stress tensor for gravitational waves  $
T_{\mu\nu}^{gw}$ discussed in Appendix \ref{gw} (see Eq. (\ref{Tmunu.gw})).   \\ \noindent
Since gravity couples to the energy momentum tensor, it is clear
that the spacetime geometry (i.e. the gravitational field) will be
affected by the presence of quantum matter fields. \\ \noindent
The secular part of the backreaction
of quantum fields on the spacetime
geometry is described by the semiclassical Einstein equations \be
\label{backeqs} R_{\mu\nu}-\frac{1}{2}Rg_{\mu\nu}=8\pi \langle
T_{\mu\nu}\rangle \ , \ee where the r.h.s. should be computed, in
the case of our interest, for an arbitrary spherically symmetric
black hole geometry. The actual black hole metric is obtained by
solving Eqs. (\ref{backeqs}) self consistently for $g_{\mu\nu}$.
Up to now the explicit form of $\langle T_{\mu\nu} (g)\rangle$ is
not known in four dimensions; this makes it impossible to solve
Eqs. (\ref{backeqs}). \\
Taking into account the vacuum contribution
of the quantum matter fields, Schwarzschild spacetime is not a
solution of the semiclassical equations (\ref{backeqs}) being the
l.h.s. identically zero whereas $\langle T_{\mu\nu}
(g^{schw})\rangle \neq 0$ (see for instance the analytical approximations
developed in \cite{anderson, analapp}). \\
\noindent A black hole dressed by the quantum contribution of the 
energy momentum tensor 
can be
properly described by 
the
metric \cite{bardeen, York}
\be \label{backmetric}
ds^2=-e^{2\psi(r,v)}(1-2m(v,r)/r)dv^2+2e^{\psi(r,v)}dvdr +
r^2d\Omega^2\ ,\ee where the functions $\psi$ and $m$ are
determined by Eqs. (\ref{backeqs}), namely  \bea
\frac{\partial m}{\partial r}&=& 4\pi r^2 \langle T^v_{\ v}\rangle \ , \nonumber \\
\frac{\partial m}{\partial v}&=& 4\pi r^2 \langle T^r_{\ v}\rangle \ , \nonumber \\
\frac{\partial \psi}{\partial r}&=&  4\pi r\langle T_{rr}\rangle =
4\pi r e^{\psi}\langle T^v_{\ r}\rangle  \ . \label{backeqs3}\eea 
To describe an evaporating black hole, the
expectation values on the r.h.s. of Eqs. (\ref{backeqs3}) are to
be taken in the $in$ vacuum state. 
(A numerical integration of these equations has been carried
out in \cite{PP}. It gives the entire space-time
from the collapse till the end of the evaporation process.
To obtain a closed set of equations,
the mean flux $\langle T_{\mu\nu}\rangle$ has been taken to be that
of a two-dimensional field divided by $4\pi r^2$. An analytic
self-consistent calculation of the same flux has been
performed in \cite{Mass} in the adiabatic regime $dM/dv \ll M^2$. It confirms
that the calculation of the Bogoliubov coefficients made in the
static geometry offers a good approximation as long as the above
adiabatic condition is satisfied.) 
\\ \noindent The apparent
horizon of an evaporating black hole is given by \be
r_h=r(v,r=2m)\equiv 2M(v)\ .\ee Near this horizon  the leading
behavior of $\langle T_{\mu\nu}\rangle$ represents an ingoing
flux of negative energy radiation going down the hole \cite{candelas}. \\
\noindent For $r\sim r_h$ the evaporating black hole metric
(\ref{backmetric}) can be therefore approximated as an ingoing
Vaidya metric \be ds^2=-(1-2M(v)/r)dv^2 + 2dvdr + r^2d\Omega^2\ee
and the relevant backreaction equation reads \be
\frac{dM}{dv}=4\pi r^2\langle T^r_{\ v}\rangle \ee giving the rate
at which the horizon shrinks in size. By energy conservation, at
infinity this should equal the luminosity $L$ radiated by
the hole. In the case of thermal radiation, $L\sim A T^4$, where $A$
is roughly the horizon area. \\ \noindent For a Schwarzschild
black hole of mass $M_0$, this yields \be L\sim AT_H^4 \sim M_0^2
k^4 \sim 1/M_0^2 \ . \ee Extrapolating this result in the case of
slow variation, one expects the mass loss of the system at
infinity to be given by \be \frac{dM}{dt}\sim -\frac{1}{M^2}\ .
\ee As the mass decreases because of Hawking radiation the process
accelerates leading to a time scale for the complete evaporation
as \be t \sim M_0^{3} \ . \ee For black holes with a conserved
charge
(Reissner-Nordstr\"om) the evolution appears quite different. \\
\noindent Arguments similar to the one given for a Schwarzschild
black hole lead to approximate the charged evaporating black hole
metric near the horizon with a charged incoming Vaidya metric \be
ds^2=-(1-2M(v)/r +Q^2/r^2)dv^2 + 2dvdr + r^2d\Omega^2 \ . \ee The
asymptotic mass loss is, as before, expected to be \be
\frac{dM}{dt}\sim - AT_H^4 \sim -r_+^2 k_{rn}^4 \ee where $k_{rn}$
and $r_+$ are given, respectively, in Eqs. (\ref{horrn}) and
(\ref{horrntwo}).
Suppose one starts from a black hole with $M\gg |Q|$. The Hawking
temperature increases as the hole looses mass, like in the
Schwarzschild case. However, when $M$ drops to
$\frac{2}{\sqrt{3}}|Q|$ one enters a completely different regime
in which $k_{rn}$ and hence $T_H$ decrease as the mass goes down,
approaching zero when $M=|Q|$. So the end state of Hawking
evaporation for a black hole with a conserved U(1) charge is an
extremal black hole, which is reached asymptotically in time as
one can see as follows. For a near extremal black hole ($M
\stackrel{>}{\sim} |Q|$) we can write  \be \frac{dM}{dt}\sim -
AT_H^4 \sim - M^2 \frac{(M-|Q|)^2M^2}{M^8}\ . \ee Integration
gives \be t\sim \frac{1}{M-|Q|}\ . \ee

\section{Quantum stress tensor and vacuum states in 2 dimensions}
\label{QuantumT} In order to consider backreaction effects due to
Hawking radiation in acoustic black holes we need to construct the
quantum stress $\langle T_{\mu\nu}\rangle$ of the phonons (scalar)
fields  $\psi_1$.  Starting with  the action (\ref{S2}), we
consider the near-horizon approximation  which is equivalent to
have a free scalar field propagating in the two-dimensional radial
part of the spacetime metric. Our reduced scalar field is then
described by the classical action \be S=-\frac{1}{2}\int
d^2x\sqrt{-g}(\nabla \phi)^2 \ \ee with stress tensor \be
T_{\mu\nu}=-\frac{2}{\sqrt{-g}}\frac{\delta S}{\delta
g_{\mu\nu}}=\partial_{\mu}\phi\partial_{\nu}\phi -
\frac{1}{2}g_{\mu\nu}(\nabla \phi)^2 \ \ . \ee The tracelessness
property $T^{\mu}_{\mu}=0$ is a consequence of the conformal
invariance of the classical action, whereas the conservation
$\nabla^{\mu}T_{\mu\nu}=0$ derives from diffeomorphism invariance.
Unfortunately these two properties cannot hold simultaneously in
the quantum theory (see for instance \cite{bd, fabnav}). If, as
usual, we choose to maintain the latter the covariant quantum
stress tensor for the scalar fields acquires an anomalous trace
which is entirely geometrical and state independent \be
\label{trace}\langle T^{\mu}_{\mu} \rangle = \frac{\hbar
R}{24\pi}\ , \ee
where $R$ is the Ricci scalar. Choosing a conformal frame for the 2d metric
\be\label{confframe} ds^2=-Cdx^+dx^-=e^{2\rho}dx^+dx^- \ ,\ee Eq. (\ref{trace}) is
translated into \be\label{t+-} \langle T_{+-} \rangle =-\frac{\hbar}{12\pi}\partial_+\partial_-\rho \ . \ee The remaining
components $\langle T_{\pm\pm}\rangle$ can be (almost completely)
determined by imposing the conservation equations \be
\partial_{\mp}\langle T_{\pm\pm}\rangle +
\partial_{\pm}\langle T_{+-}\rangle - 2\partial_{\pm}\rho\langle
T_{+-}\rangle=0\ .\ee The result is \be\label{t++--} \langle T
_{\pm \pm} \rangle =-\frac{\hbar}{12\pi}
C^{1/2}(C^{-1/2})_{,\pm\pm}+t_{\pm}(x^{\pm}) . \ee The ambiguity
that remains are the two arbitrary functions $t_{\pm}(x^{\pm})$,
which encode the information about the quantum state in which the
expectation values are taken.

In a generic conformal coordinate system $x^{\pm}$ (Eq.
(\ref{confframe})) we can define a quantization associated to the
choice of the positive frequency\index{positive frequency} modes
\be \label{modesx} (4\pi w)^{-1/2}e^{-iwx^+}, \ \ \ (4\pi
w)^{-1/2}e^{-iwx^-} . \ee Then, if we choose the state $|\
\rangle$ to be the vacuum state with respect to the modes
(\ref{modesx}) (usually denoted as $|x^{\pm}\rangle$) the
functions $t_{\pm}$ vanish and we get
 \be\label{definitionqst5}
\langle x^{\pm}|T_{\pm\pm}(x^{\pm})|x^{\pm}\rangle =-\frac{\hbar}{12\pi}(\partial_{\pm}\rho\partial_{\pm}\rho -
\partial_{\pm}^2\rho)\ .
\ee Moreover,  a change of vacuum state $|x^{\pm}\rangle
\rightarrow|\tilde x^{\pm} \rangle$ produces a change in the
expectation values of the $T_{\pm\pm}(x^{\pm})$ components of the
stress tensor of the form \be\label{changevacuum} \langle \tilde
x^{\pm}| T_{\pm\pm}|\tilde x^{\pm} \rangle = \langle x^{\pm}|
T_{\pm\pm}| x^{\pm} \rangle - \frac{\hbar}{24\pi}\{\tilde
x^{\pm},x^{\pm}\}\ , \ee where $\{\tilde x^{\pm},x^{\pm}\}$ is the
Schwarzian derivative \be\label{schwder} \{ \tilde x^{\pm},
x^{\pm}\}= \frac{d^3 \tilde x^{\pm}}{dx^{\pm 3}}/ \frac{d\tilde
x^{\pm }}{dx^{\pm}} -\frac{3}{2} \left(\frac{d^2 \tilde
x^{\pm}}{dx^{\pm 2}}/\frac{d\tilde x^{\pm }}{dx^{\pm}}\right)^2 \
. \ee \\ \noindent For an acoustic geometry
$ds^2=-\frac{\rho(c^2-v^2)}{c^3}dx^+dx^-$, where \bea
x^+&=&c\left(t+\int \frac{dz}{c-v}\right)\ , \\
x^-&=&c\left(t-\int \frac{dz}{c+v}\right)\ , \eea ingoing modes
are positive frequency with respect to $t$, whereas near the
horizon outgoing modes are positive frequency with respect to 
$X^-=-\k^{-1}e^{-\k x^-}$. So, evaluating the Schwarzian derivative, we
have $t_+=~0,\, t_- =\frac{\hbar \k^2}{48 \pi }$ 
and the stress tensor reads \cite{lungo}
\bea \langle
T_{++}\rangle&=&-\frac{\hbar}{12\pi}\frac{c^2-v^2}{4c^4}
\left[\left(-\frac{\rho''}{2\rho}+\frac{3\rho'^2}{4\rho^2} \right)
(c^2-v^2)
+v'^2+vv''+\frac{(vv')^2}{c^2-v^2} \right]\ , \ \ \ \ \ \ \ \ \\
\langle T_{--}\rangle&=&\langle T_{++}\rangle+\frac{\hbar \k^2}{48\pi}\ ,\\
\langle T\rangle&=&-\frac{\hbar}{24\pi\rho c} \left[ \left(
\frac{\rho''}{\rho}-\frac{\rho'^2}{\rho^2} \right)(c^2-v^2)-
\left( \frac{2vv'\rho'}{\rho}+2v'^2+2vv''\right) \right] \ .\ \ \
\ \ \eea

\end{appendix}


\end{document}